\documentclass[11pt]{article}
\usepackage[margin=1in]{geometry}
\usepackage[T1]{fontenc}
\usepackage{lmodern,mathtools,amsmath,amssymb,amsthm,booktabs,tabularx,longtable,graphicx,microtype,caption,float,enumitem,placeins,needspace}
\usepackage{subfiles}
\usepackage[colorlinks=true,linkcolor=blue,urlcolor=blue]{hyperref}
\newcolumntype{L}{>{\raggedright\arraybackslash}X}
\usepackage{tikz}
\newtheorem{lemma}{Lemma}
\newtheorem{theorem}{Theorem}
\newtheorem{proposition}{Proposition}
\newtheorem{corollary}{Corollary}
\theoremstyle{remark}

\theoremstyle{plain}
\newcommand{\E}{\mathbb E}
\newcommand{\Pp}{\mathbb P}
\newcommand{\Prb}{\mathbb P}
\newcommand{\diam}{\operatorname{diam}}
\newcommand{\TV}{\operatorname{TV}}
\newcommand{\KL}{\operatorname{KL}}
\newcommand{\dist}{\operatorname{dist}}
\newcommand{\src}[1]{\csname cite#1\endcsname}

\title{Honest confidence sets under experiment-level failure}
\author{
Yuxin Han\textsuperscript{1},
Lanjun Cheng\textsuperscript{2},
Fengshi Jing\textsuperscript{1}\thanks{Corresponding author: Fengshi Jing.}\\[0.6em]
\small \textsuperscript{1}Faculty of Data Science,
City University of Macau, Macao SAR, China\\
\small \textsuperscript{2}Franklin College of Arts and Sciences,
University of Georgia, Athens, GA, USA
}
\date{}
\begin{document}
\maketitle

\begin{abstract}
An experiment-level failure replaces the joint observation law while preserving the target. We study honest confidence sets when a declared upper bound on failure probability equals the allowed noncoverage probability. For a Uniform support endpoint and a Gaussian mean on fixed compact parameter intervals, every honest deterministic procedure has full-range worst-case expected diameter under arbitrary or absolutely continuous joint failures. Independent randomisation gives sharp minimax risks of $(2+o(1))\log N/N$ for the endpoint in $[1,2]$ and $(2\sigma+o(1))\sqrt{\log N/N}$ for the mean with known variance $\sigma^2$. A failure distribution shared by many targets forces average clean-state noncoverage to decrease with expected diameter. One-sided local Kullback--Leibler bounds then yield lower bounds with logarithmic factors. The endpoint experiment also has a sharp deterministic frontier as the remaining error budget vanishes. Under exchangeable failures without ties, observation order recovers both randomised benchmark rates deterministically. In a reference-calibrated roster, we derive the randomised minimax reliability rate. A prespecified reference reserve of logarithmic size permits deterministic attainment of the same order on a subclass whose conditionally independent reserved responses have probabilities bounded away from zero and one in every state.
\end{abstract}

\section{Introduction}

An entire statistical experiment may fail even when its clean-state model is well understood. Suppose a 5\% upper bound on failure probability is available and a confidence procedure must attain 95\% coverage uniformly over the allowed failures. How much precision remains when the failure bound exhausts the noncoverage budget? We model the joint observation law as $P_{\theta,\lambda,Q}=(1-\lambda)G_{\theta,N}+\lambda Q$, where $G_{\theta,N}$ is the clean law and $Q$ is a failure law for the complete observation vector. The target $\theta$ is fixed across states. The actual failure probability $\lambda$ can take any value in $[0,\eta]$, where $\eta$ is a declared upper bound. Coverage and expected-diameter risk are evaluated over this same class of mixture laws. We focus on the boundary $\eta=\alpha$, where $\alpha$ is the allowed noncoverage probability.

This model is a Huber contamination neighbourhood on the sample space of the entire experiment~\cite{huber1964robust}. Huber's robust confidence limits~\cite{huber1968robust} already use randomised minimax procedures. Recent work studies adaptive robust confidence intervals and sets under observation-wise contamination for Gaussian means~\cite{luo2026adaptive}, binomial proportions~\cite{cho2025robust}, linear regression~\cite{xie2026confidence} and high-dimensional Gaussian mean vectors~\cite{diakonikolas2026adaptive}. Related work considers contamination introduced before Gaussian measurement noise~\cite{wang2026adaptive}. Here the failure law acts on the full joint experiment, and the critical bound exhausts the noncoverage budget. We quantify the resulting minimax separation between deterministic and independently randomised procedures and identify two settings in which usable randomisation can be recovered from the observations.

At the boundary, every honest deterministic rule for a Uniform support endpoint or a Gaussian mean has full-range worst-case expected diameter under arbitrary or absolutely continuous joint failures. Independent randomisation permits shrinking diameter in both experiments. For an endpoint in $[1,2]$, the sharp randomised minimax risk is $(2+o(1))\log N/N$. For a Gaussian mean on a fixed compact interval with known variance $\sigma^2$, it is $(2\sigma+o(1))\sqrt{\log N/N}$. These rates inflate the clean scales $N^{-1}$ and $N^{-1/2}$ by logarithmic and square-root-logarithmic factors, respectively. The separation arises because a failure law can make a deterministic rule's clean error event certain, while an independent seed remains valid under every allowed joint failure.

A failure distribution shared by many targets explains the information cost. Under that law, a short confidence set covers only a small fraction of the parameter range. Uniform coverage therefore forces average clean-state noncoverage to decrease with expected diameter. Combined with a uniform one-sided local Kullback--Leibler upper bound of order $Nh^\kappa$, this coverage budget yields the lower rate $(\log N/N)^{1/\kappa}$ for interval parameters and sufficiently fine grids (Corollary~\ref{cor:gp-rate}). Exact support and Gaussian testing comparisons, together with matching constructions, determine the benchmark constants. The argument connects expected confidence-set length to testing in honest adaptation~\cite{cai2004adaptation} and to model-class separation in impossibility theory~\cite{bertanha2020impossible}, with the additional coverage constraint imposed by a common failure law.

The deterministic frontier in the endpoint experiment describes how precision changes as the failure bound approaches the allowed noncoverage probability (Theorem~\ref{slack:sharp}). Polynomial slack permits shrinking diameter, exponentially small slack gives a positive limiting diameter, and zero slack forces full-range risk. At the boundary, exchangeability without ties preserves enough randomness in observation order to recover both benchmark rates deterministically. Permutation-invariant deterministic rules retain full-range risk. Randomised intervals for discrete models~\cite{geyer2005fuzzy,kabaila2013randomized}, data-randomised intervals~\cite{korn1987data}, and rank-based randomisation~\cite{ramdas2026randomized} provide related constructions. Our results quantify the minimax consequences of requiring the random source to remain valid under every allowed failure law.

Population-count inference in a fixed roster gives a further setting in which calibration and usable randomness interact. Membership in the main roster is latent, and a separate roster of known members calibrates a binary response mechanism. The minimax reliability rate separates main-sample noise, reference-sample noise, calibration tolerance and background ambiguity. Within a subclass whose reserved reference responses are conditionally independent and have probabilities bounded away from zero and one in every state, a prespecified reserve of logarithmic size supplies an internal seed and attains the randomised rate deterministically. The reference sample thus has two statistical roles, providing calibration information and preserving randomness under experiment-level failure.

\section{Benchmark experiments at the failure boundary}\label{ep:section}

Let $\theta$ belong to a fixed compact interval $I$ of length $D>0$. The observation is an indexed vector with law
\begin{equation}\label{ep:model}
P_{\theta,\lambda,Q}=(1-\lambda)G_{\theta,N}+\lambda Q,
\qquad 0\le\lambda\le\eta.
\end{equation}
The parameter is the same in both states. The failure law $Q$ may depend on $\theta$ and $N$. We first set the declared bound to $\eta=\alpha$, where $0<\alpha<1/2$ is fixed, while allowing every actual failure probability $0\le\lambda\le\alpha$. A confidence set $C(X,\xi)\subseteq I$ may use a seed $\xi$ independent of the state and observations, with a common law throughout the model. It is honest if
\begin{equation}\label{ep:honesty}
\sup_{\theta\in I,\,0\le\lambda\le\alpha,\,Q\in\mathcal Q_N}
\Prb_{P_{\theta,\lambda,Q},\xi}\{\theta\notin C\}\le\alpha.
\end{equation}
Coverage averages over the experiment state and any auxiliary seed under each allowed mixture law. Write $R_N(C;\mathcal Q_N)$ for the same supremum of $\E\diam C$, and $\mathcal R^{\mathrm{rand}}_{N,\mathcal Q}$ for its infimum over honest procedures. The deterministic risk $\mathcal R^{\mathrm{det}}_{N,\mathcal Q}$ restricts this infimum to functions of $X$ alone. A full-range output has finite diameter $D$ on the compact parameter interval and contributes to expected-diameter risk in proportion to its probability. Sets have measurable inclusion graphs and diameters, with $\diam\varnothing=0$.

On each experiment's sample space, let $\mathcal Q_N^{\mathrm{all}}$ contain all probability laws, $\mathcal Q_N^{\mathrm{ac}}$ all laws absolutely continuous with respect to Lebesgue measure, and $\mathcal Q_N^{\mathrm{ex}}$ all exchangeable laws without ties. Deterministic procedures may use the original coordinate indices. A permutation-invariant procedure additionally satisfies $C(\pi x)=C(x)$ for every coordinate permutation $\pi$.

\subsection{A Uniform support endpoint}

For $\theta\in[1,2]$, let $G_{\theta,N}=\operatorname{Unif}[0,\theta]^{\otimes N}$, with failure laws on $[0,2]^N$.
\begin{theorem}[Sharp endpoint risk]\label{ep:sharp}
For every $N\ge1$ and $\mathcal Q\in\{\mathcal Q^{\mathrm{all}},\mathcal Q^{\mathrm{ac}}\}$,
\begin{equation}\label{ep:det}
\mathcal R^{\mathrm{det}}_{N,\mathcal Q}=1.
\end{equation}
For each of the three failure classes,
\begin{equation}\label{ep:sharpconstant}
\mathcal R^{\mathrm{rand}}_{N,\mathcal Q}
=(2+o(1))\frac{\log N}{N}.
\end{equation}
An attaining construction uses the maximum $M_N=\max_iX_i$ and
\begin{equation}\label{ep:base}
B_\beta(M_N)=[\max\{1,M_N\},\min\{2,M_N/\beta^{1/N}\}],
\end{equation}
with reversed intervals interpreted as empty. Independently return $[1,2]$ with probability
\begin{equation}\label{ep:fallback}
p_\beta=\frac{(1-\alpha)\beta}{\alpha+(1-\alpha)\beta},
\end{equation}
and otherwise return $B_\beta$. For $0<\beta\le1$, the risk is at most
\begin{equation}\label{ep:finiteupper}
p_\beta+(1-p_\beta)D_{N,\beta},\qquad
D_{N,\beta}=\min\{1,2(1-\beta^{1/N})\}.
\end{equation}
The choice $\beta=1/N$ attains~\eqref{ep:sharpconstant}.
\end{theorem}

\noindent\textbf{Remark (parameter scaling).} The leading constant $2$ reflects the upper endpoint of the parameter range. For a fixed interval $[\ell,u]\subset(0,\infty)$ with $\ell<u$, the same argument gives $(u+o(1))\log N/N$.

\noindent\textbf{Deterministic obstruction.} Fix a deterministic rule. If its clean error at $\theta$ is positive, condition $G_{\theta,N}$ on that error event. The resulting failure law is absolutely continuous and makes the rule miss $\theta$ with probability one. At $\lambda=\alpha$, the mixture error then exceeds $\alpha$. Clean coverage must therefore be perfect. Since $G_{1,N}$ is absolutely continuous with respect to $G_{2,N}$, the rule must contain both endpoints almost surely under $G_{1,N}$, giving diameter one.

\noindent\textbf{Randomised construction and sharpness.} The independent seed remains outside the failure law's control. The base interval has clean error at most $\beta$ and pointwise diameter at most $D_{N,\beta}$. Its mixture error is at most $\alpha+(1-\alpha)\beta$, so the independent full-range decision reduces the error to $\alpha$. The matching lower bound combines the coverage budget in Section~\ref{sec:general-principle} with the likelihood ratio between nested Uniform supports (Supplement Section~\ref{supp-sec:endpoint-proofs}).

\begin{proposition}[Recovery from observation order]\label{ep:exchangeable}
For exchangeable failure laws without ties,
\begin{equation}\label{ep:exchangeableconstant}
\mathcal R^{\mathrm{det}}_{N,\mathcal Q^{\mathrm{ex}}}
=(2+o(1))\frac{\log N}{N}.
\end{equation}
Let $J=\arg\max_iX_i$. Return $[1,2]$ when $J\le\lceil Np_\beta\rceil$ and $B_\beta(M_N)$ otherwise. This deterministic rule is honest and has risk at most
\begin{equation}\label{ep:internalfinite}
p_\beta+(1-p_\beta)D_{N,\beta}+N^{-1}.
\end{equation}
\end{proposition}
Under every exchangeable law without ties, $J$ is uniform and independent of the maximum. Its common uniform law preserves this independence after mixing the clean and failure states. The original index therefore supplies the correction coin, with rounding cost at most $1/N$.

\begin{corollary}[The cost of permutation invariance]\label{ep:invariant}
Under $\mathcal Q_N^{\mathrm{ex}}$, permutation-invariant deterministic procedures have minimax risk one for every $N\ge1$.
\end{corollary}
Conditioning a clean law on an invariant error event preserves exchangeability and the absence of ties. The preceding deterministic argument therefore applies to this procedure class as well.

\subsection{A Gaussian mean}\label{gaussian:section}

The same boundary changes the confidence scale in a regular experiment. Fix a compact interval $I=[a,b]$ of length $D=b-a>0$ and a known $\sigma>0$, and let
\begin{equation}\label{gaussian:model}
G_{\theta,N}=\mathcal N(\theta,\sigma^2)^{\otimes N},\qquad
P_{\theta,\lambda,Q}=(1-\lambda)G_{\theta,N}+\lambda Q,
\quad \theta\in I,\quad 0\le\lambda\le\alpha.
\end{equation}
The three failure classes now comprise all laws, absolutely continuous laws, and exchangeable laws without ties on $\mathbb R^N$. Use the same honesty and worst-case expected-diameter definitions as above, with outputs in $I$, and denote the Gaussian minimax risks by $\mathcal R^{\mathrm{rand},G}_{N,\mathcal Q}$ and $\mathcal R^{\mathrm{det},G}_{N,\mathcal Q}$. The constants $I$, $\sigma$ and $\alpha\in(0,1/2)$ are fixed as $N$ grows.

\begin{theorem}[Sharp Gaussian risk]\label{gaussian:sharp}
For each of the three failure classes,
\begin{equation}\label{gaussian:sharpconstant}
\mathcal R^{\mathrm{rand},G}_{N,\mathcal Q}
  =(2\sigma+o(1))\sqrt{\frac{\log N}{N}}.
\end{equation}
For arbitrary or absolutely continuous failures, $\mathcal R^{\mathrm{det},G}_{N,\mathcal Q}=D$ for every $N$. For exchangeable failures without ties,
\begin{equation}\label{gaussian:exchangeableconstant}
\mathcal R^{\mathrm{det},G}_{N,\mathcal Q^{\mathrm{ex}}}
  =(2\sigma+o(1))\sqrt{\frac{\log N}{N}},
\qquad
\mathcal R^{\mathrm{det,inv},G}_{N,\mathcal Q^{\mathrm{ex}}}=D.
\end{equation}
The last risk restricts deterministic rules to be permutation invariant.
\end{theorem}

\noindent\textbf{Construction and recovery from observation order.} An attaining construction starts with the usual Gaussian interval, clipped to the parameter range,
\begin{equation}\label{gaussian:base}
B_\beta(\bar X)=I\cap
\left[\bar X-\frac{\sigma z_\beta}{\sqrt N},
      \bar X+\frac{\sigma z_\beta}{\sqrt N}\right],
\qquad z_\beta=\Phi^{-1}(1-\beta/2).
\end{equation}
It has clean noncoverage at most $\beta$ and diameter at most $\min\{D,2\sigma z_\beta/\sqrt N\}$ at every observation vector. Return $I$ with independent probability $p_\beta$ from \eqref{ep:fallback}, and $B_\beta$ otherwise. Taking $\beta=N^{-1/2}$ yields~\eqref{gaussian:sharpconstant}. For exchangeable data without ties, the original index of the largest observation is uniform and independent of $\bar X$ under every allowed law. Using that index to implement the full-range decision, with probability rounded upward to a multiple of $1/N$, costs at most $D/N$.

\noindent\textbf{Sharp lower bound and deterministic obstruction.} One Gaussian law can serve as the failure law for every parameter. Honesty then bounds the total clean error over $I$ by a constant times the expected diameter. A small diameter therefore supplies two nearby parameters with very small clean errors. The Gaussian testing curve transfers both coverage statements to their midpoint, forcing the set to span a distance approaching $2\sigma\sqrt{\log N/N}$. For a deterministic rule, conditioning a clean law on any error event makes that event certain under an allowed failure law. Honesty forces zero clean error, and the common support of Gaussian laws then forces full-range diameter. The complete proof and finite-sample bounds are given in Supplement Section~\ref{supp-gaussian:proofs}.

\section{Coverage budgets and information limits}\label{sec:general-principle}

The two benchmark lower bounds share a coverage constraint. Under a failure distribution compatible with many targets, a short confidence set has small average coverage over those targets. Honesty at the boundary then forces small average clean-state error (Lemma~\ref{lem:gp-budget}). Distinguishing nearby clean laws at these error levels requires a wider confidence set.

Consider a fixed interval $I$ of length $L>0$, with parameter set $\Theta_N=I$ or a nonempty regular grid $I\cap(u_N+\Delta_N\mathbb Z)$. Each model class $\mathcal P_{\theta,N}$ contains $G_{\theta,N}$ and $(1-\alpha)G_{\theta,N}+\alpha Q_N$, where $Q_N$ is common to all targets. With the preceding measurability and seed conventions, put
\begin{equation}\label{eq:gp-risk}
R=\sup_{\theta\in\Theta_N,\,P\in\mathcal P_{\theta,N}}
\E_{P,\xi}\diam C,\qquad
e_\theta=\Prb_{G_{\theta,N},\xi}\{\theta\notin C\}.
\end{equation}
For interval parameters, the clean laws form a measurable probability kernel. Honesty is uniform over $\theta$ and $P$.

\begin{lemma}[Coverage budget]\label{lem:gp-budget}
Put $c_\alpha=\alpha/(1-\alpha)$. Every honest procedure satisfies
\begin{align}
\int_Ie_\theta\,d\theta&\le R/(1-\alpha),
&&\Theta_N=I,\label{eq:gp-budget-continuous}\\
\Delta_N\sum_{\theta\in\Theta_N}e_\theta
&\le R/(1-\alpha)+c_\alpha\Delta_N,
&&\Theta_N\text{ a grid}.\label{eq:gp-budget-grid}
\end{align}
If $Q_N$ is itself an allowed observation law at some parameter, the right sides improve to $c_\alpha R$ and $c_\alpha(R+\Delta_N)$, respectively.
\end{lemma}
Indeed, coverage under the critical mixture gives
\begin{equation}\label{eq:gp-pointwise-budget}
(1-\alpha)e_\theta\le\alpha\Prb_{Q_N,\xi}\{\theta\in C\}.
\end{equation}
Integrate over the targets. The length covered by a set is at most its diameter. On a grid, its covered mass is at most the diameter plus one mesh width. These inequalities give the coverage budget.

\begin{corollary}[Logarithmic information cost]\label{cor:gp-rate}
Fix $\kappa,C_0,h_0>0$ and an interval $J$ of positive length with $J+[0,h_0]\subseteq I$. Suppose, uniformly for all sufficiently large $N$,
\begin{equation}\label{eq:gp-local-kl}
\KL(G_{\theta,N}\Vert G_{\theta+h,N})\le C_0Nh^\kappa
\end{equation}
for every $\theta\in J$ and $0<h\le h_0$ in the interval case, or for every $\theta\in J\cap\Theta_N$ and $h\in\Delta_N\mathbb N\cap(0,h_0]$ in the grid case. For grids assume
\begin{equation}\label{eq:gp-mesh}
\Delta_N=o\bigl((\log N/N)^{1/\kappa}\bigr).
\end{equation}
The minimax risk, allowing randomisation, obeys
\begin{equation}\label{eq:gp-rate}
\mathcal R_N^*\ge c(\log N/N)^{1/\kappa}
\end{equation}
for all sufficiently large $N$, where $c>0$ depends only on the fixed constants.
\end{corollary}
\noindent\textbf{Proof mechanism.} A short set under one clean law excludes a sufficiently separated target with high probability. The KL bound transfers this exclusion event to the neighbouring law and bounds its error from below. Integrating these errors and applying Lemma~\ref{lem:gp-budget} gives~\eqref{eq:gp-rate}. The finite bound compares a separation term of order $d$ with an exponential term of order $\exp(-CNd^\kappa)$, with an $O(\Delta_N)$ correction on grids. Choosing $d$ as a sufficiently small fixed multiple of $(\log N/N)^{1/\kappa}$ keeps the exponential term above a constant multiple of $d$ and makes the mesh correction $o(d)$. Supplement Sections~\ref{supp-sec:general-finite}--\ref{supp-sec:gp-rate-proof} give the finite inequality and full proof.

For the Uniform endpoint, $\KL(G_{\theta,N}\Vert G_{\theta+h,N})=N\log(1+h/\theta)\le Nh$. For the Gaussian mean, the divergence is $Nh^2/(2\sigma^2)$. Corollary~\ref{cor:gp-rate} therefore gives the logarithmic and square-root-logarithmic lower rates. The sharp benchmark constants follow from the exact testing comparisons and matching constructions in Section~\ref{ep:section}.
\label{sec:upper-transfer}

The lower bound is complemented by a construction whose diameter is controlled in every state. Let $\Theta$ be a compact interval or a finite grid with diameter $D>0$. At target $\theta$, consider laws $(1-\lambda)G+\lambda Q$, where $0\le\lambda\le\alpha$, $G$ is an allowed clean law and $Q$ is an allowed failure law. Either component can depend on $\theta$.

\begin{proposition}[Independent full-range correction]
\label{prop:upper-transfer}
Let $0<\beta\le1$. Suppose a measurable set $B_\beta(X)\subseteq\Theta$ satisfies
\[
\sup_{\theta,G}\Prb_G\{\theta\notin B_\beta\}\le\beta,
\qquad
\max\left\{\sup_{\theta,G}\E_G\diam B_\beta,
           \sup_{\theta,Q}\E_Q\diam B_\beta\right\}\le w_N(\beta).
\]
Here $0\le w_N(\beta)\le D$. Use an independent seed $\xi\sim\operatorname{Unif}[0,1]$ and return $\Theta$ when $\xi\le p_\beta$, where
\[
p_\beta=\frac{(1-\alpha)\beta}{\alpha+(1-\alpha)\beta}.
\]
Return $B_\beta$ otherwise. The resulting procedure is honest over all the specified mixtures and has worst-case expected diameter at most
\begin{equation}\label{eq:upper-transfer}
p_\beta D+(1-p_\beta)w_N(\beta).
\end{equation}
\end{proposition}

The coverage claim follows from $(1-p_\beta)\{\alpha+(1-\alpha)\beta\}=\alpha$. Conditioning on the full-range decision gives the diameter bound.

A pointwise bound $\diam B_\beta(x)\le w_N(\beta)$ is sufficient for the two expectation bounds. The endpoint construction has such a bound. In the roster experiment, the reference restrictions give a uniform expectation bound in every state.

Replacing an empty output by a prespecified singleton in $\Theta$ preserves its diameter and weakly increases coverage. The same risk bounds are therefore attainable with nonempty outputs.

\section{The Uniform endpoint error-budget frontier}\label{slack:section}

The boundary results leave open how precision changes when some noncoverage budget remains after allowing for failure. Return to the endpoint experiment with $\theta\in[1,2]$. Let the failure probability satisfy $\lambda\le\eta<1$ and let $\bar\alpha\in(0,1)$ be the allowed total noncoverage. Write $\mathcal R_N^{\mathrm{det}}(\eta,\bar\alpha;\mathcal Q)$ for the minimax expected diameter over the corresponding mixture class. Consider arbitrary deterministic rules under $\mathcal Q^{\mathrm{all}}$ or $\mathcal Q^{\mathrm{ac}}$, or permutation-invariant deterministic rules under $\mathcal Q^{\mathrm{ex}}$.

\begin{theorem}[Sharp deterministic error-budget frontier]
\label{slack:sharp}
In each of these settings, if $\eta\ge\bar\alpha$, then
\begin{equation}\label{slack:exhausted}
\mathcal R_N^{\mathrm{det}}(\eta,\bar\alpha;\mathcal Q)=1
\qquad(N\ge1).
\end{equation}
If $\eta<\bar\alpha$, define
\begin{equation}\label{slack:definitions}
\beta=\frac{\bar\alpha-\eta}{1-\eta},\qquad
t=\log(1/\beta),\qquad
D_N(t)=\min\{1,2(1-e^{-t/N})\}.
\end{equation}
For every $0<s<t$,
\begin{equation}\label{slack:finite}
D_N(t-s)[1-e^{-t}-e^{-s}]_+
\le\mathcal R_N^{\mathrm{det}}(\eta,\bar\alpha;\mathcal Q)
\le D_N(t).
\end{equation}
Consequently, whenever $\beta_N\to0$,
\begin{equation}\label{slack:asymptotic}
\mathcal R_N^{\mathrm{det}}(\eta_N,\bar\alpha_N;\mathcal Q)
=(1+o(1))\min\{1,2(1-\beta_N^{1/N})\}.
\end{equation}
The relative approximation holds for every relationship between $\log(1/\beta_N)$ and $N$.
\end{theorem}

A failure law can concentrate on a deterministic rule's clean error event. Thus robust honesty with $\eta<\bar\alpha$ is equivalent to clean noncoverage at most $\beta=(\bar\alpha-\eta)/(1-\eta)$. The interval $B_\beta$ in~\eqref{ep:base} meets that requirement and has diameter at most $D_N(t)$ for every observation. Conversely, nested Uniform supports transfer coverage of the upper endpoint to a smaller parameter. Under that smaller parameter, both values must be covered with high probability, forcing almost the same diameter. Taking $s=\sqrt t$ in~\eqref{slack:finite} gives the relative approximation as $t\to\infty$ (Supplement Section~\ref{supp-slack:proofs}).

For example, keep the failure bound at $\eta=\alpha$ and allow noncoverage $\bar\alpha_N=\alpha+\delta_N<1$. The theorem gives
\begin{equation}\label{slack:regimes}
\begin{array}{c|c}
\text{Available error }\delta_N &
\mathcal R_N^{\mathrm{det}}(\alpha,\alpha+\delta_N;\mathcal Q)\\[2pt]
\hline
N^{-\gamma},\quad\gamma>0
 & (2\gamma+o(1))\log N/N\\[2pt]
e^{-cN},\quad c>0
 & \longrightarrow\min\{1,2(1-e^{-c})\}\\[2pt]
0 & 1
\end{array}
\end{equation}
The same regimes hold when coverage remains $1-\alpha$ and the failure bound approaches $\alpha$ with slack $\alpha-\eta_N=\delta_N$. Polynomial slack permits shrinking diameter. Exponentially small slack produces a positive limiting diameter, and exhausting the budget forces full-range risk.

\section{Reference-calibrated roster inference}\label{sec:roster}

A reference sample can both calibrate a response mechanism and supply randomness for inference. The roster experiment separates these roles while allowing heterogeneous response probabilities and a shared unobserved environment. The index-of-maximum randomisation used in Section~\ref{ep:section} is not generally uniform under heterogeneous binary response laws. The population-count target is related to prevalence correction~\cite{rogan1978estimating}, finite-population inference~\cite{ge2024enhanced,mirzahosseini2026prevalence} and partial identification with misclassified responses~\cite{molinari2008partial}.

\label{sec:roster-model}
There are $n$ main units with fixed, unknown membership labels $M_i\in\{0,1\}$ and target count $K=\sum_iM_i$. A disjoint reference roster contains $m$ known members. We observe every unit's binary response, with main total $Y=\sum_iZ_i$ and reference total $X=\sum_jV_j$. The full data retain the original indices. Given a shared environment $U$, all responses are independent Bernoulli variables with individual probabilities $q_i(U)$ and $p_j(U)$. Write
\[
r(U)=\frac1m\sum_jp_j(U),\quad
a_M(U)=\frac1K\sum_{i:M_i=1}q_i(U),\quad
b_M(U)=\frac1{n-K}\sum_{i:M_i=0}q_i(U).
\]
Calibration succeeds when
\begin{equation}\label{eq:class}
|a_M(U)-r(U)|\le\varepsilon.
\end{equation}
The reliability class $\mathcal P_\eta$ is defined by declarations fixed before the responses are observed,
\begin{equation}\label{eq:envclass}
r(U)\in[r_-,r_+],\quad b_M(U)\in[b_L,b_U]\quad\text{a.s.},\qquad
\Pp\{|a_M(U)-r(U)|>\varepsilon\}\le\eta.
\end{equation}
For inference from a single roster, $\eta$ is an externally specified bound on calibration failures across environments. Reference and background restrictions hold in every state. The member--reference calibration relation can fail. Membership is fixed across states, and individual probabilities may range over $[0,1]$ subject to these group averages. Conditioning on calibration success or failure gives the whole-experiment mixture~\eqref{ep:model}, with actual failure probability $\lambda\le\eta$.

Fix margins $0<\rho_-<\rho_+<1$, $0\le\bar b<r_-<r_+<1$ and $0\le\bar\varepsilon<r_--\bar b$. Take $0\le b_L\le b_U\le\bar b$, $0\le\varepsilon\le\bar\varepsilon$, $\delta_b=b_U-b_L$, and $K\in\mathcal K_n=\{\lceil\rho_-n\rceil,\ldots,\lfloor\rho_+n\rfloor\}$. Declarations may vary with sample size within these fixed margins. Let $D_n=\diam\mathcal K_n$. Procedures return subsets of $\mathcal K_n$. Honesty requires coverage at least $1-\alpha$ for each fixed membership vector and every law in $\mathcal P_\eta$. Write $\mathcal R^*_{\eta,\mathrm{rand}}$ and $\mathcal R^*_{\eta,\mathrm{det}}$ for the corresponding minimax expected count diameters. For the proportion target $\pi=K/n$, all diameter risks are obtained by dividing the corresponding count-scale risks by $n$.

At exact calibration and known background $b$, the familiar Rogan--Gladen identity~\cite{rogan1978estimating} is
\begin{equation}\label{eq:ratio}
\frac K n=\frac{\E(Y/n\mid U)-b}{\E(X/m\mid U)-b}.
\end{equation}
Our construction inverts the response relation directly. For a candidate $k$, put $W_k=Y-kX/m$, $J_k=[(n-k)b_L-k\varepsilon,(n-k)b_U+k\varepsilon]$, and
\[
C_H(\beta)=\left\{k\in\{0,\ldots,n\}:\dist(W_k,J_k)
\le\sqrt{\frac{n+k^2/m}{2}\log\frac2\beta}\right\}.
\]
At the true count, the conditional mean lies in $J_K$ whenever calibration succeeds. Hoeffding's inequality~\cite{hoeffding1963probability} therefore gives clean error at most $\beta$. The statewise baseline and mixture-closure results are in Supplement Section~\ref{supp-sec:roster-baseline}.

\begin{theorem}[The roster reliability rate]\label{thm:environment}
Put
\[
s=\min\{1,n^{-1/2}+m^{-1/2}\},\qquad
u=(\alpha-\eta)_+,\qquad v=(\eta-\alpha)_+.
\]
For all sufficiently large $n$, every $m\ge1$, every $\eta\in[0,1]$ and all admissible declarations,
\begin{equation}\label{eq:envrate}
\mathcal R^*_{\eta,\mathrm{rand}}\asymp
n\min\left\{1,\varepsilon+\delta_b+v+
s\sqrt{\log\frac{\mathrm e}{\min\{1,s+u\}}}\right\}.
\end{equation}
Constants depend only on the fixed margins and $\alpha$.
\end{theorem}
For fixed $\eta<\alpha$, the rate separates main noise $\sqrt n$, reference noise $n/\sqrt m$, calibration tolerance $n\varepsilon$ and background ambiguity $n\delta_b$, clipped at order $n$. At $\eta=\alpha$, the stochastic term becomes $ns\sqrt{\log(\mathrm e/s)}$. For fixed $\eta>\alpha$, the rate is of order $n$. Hence a balanced, exactly calibrated roster with known background has count risk of order $\sqrt n$, $\sqrt{n\log n}$ and $n$ for fixed $\eta<\alpha$, $\eta=\alpha$, and fixed $\eta>\alpha$, respectively. Along $n\to\infty$, relative randomised diameter vanishes exactly when $m\to\infty$, $\varepsilon,\delta_b\to0$ and $(\eta-\alpha)_+\to0$.

Let $H_\beta=C_H(\beta)\cap\mathcal K_n$, $q_\beta=\eta+(1-\eta)\beta$, and $t_\beta=(1-\alpha/q_\beta)_+$. Return $\mathcal K_n$ with independent probability $t_\beta$ and the integer hull of $H_\beta$ otherwise. The error is at most $(1-t_\beta)q_\beta\le\alpha$. Reference concentration bounds the conditional expected diameter of $H_\beta$ uniformly over states, and choosing $\beta=(s+u)/4$ attains~\eqref{eq:envrate}.

For the lower bound, a common failure law constrains clean errors across many candidate counts. Two full-data subexperiments have KL scales $(j-k)^2/n$ and $m(j-k)^2/n^2$, accounting for main and reference information. Their testing bounds combine with the coverage budget to give the logarithmic term. In the balanced critical case, Corollary~\ref{cor:gp-rate} on the proportion grid gives
\begin{equation}\label{eq:roster-critical-lower}
\mathcal R^*_{\alpha,\mathrm{rand}}/n\gtrsim\sqrt{\log n/n}.
\end{equation}
The full rate follows by combining these bounds with the calibration and background identification terms (Supplement Sections~\ref{supp-env:sec:upper}--\ref{supp-env:sec:lower}). The proportion-grid argument connects its critical logarithm to Corollary~\ref{cor:gp-rate} (Supplement Section~\ref{supp-sec:roster-grid-detail}).

\begin{theorem}[The deterministic roster obstruction]\label{thm:deterministic}
Suppose $\eta\ge\alpha$ and the background is known and rational, $b\in[0,\bar b]$. Let $d_b$ denote its denominator in lowest terms, with $d_b=1$ when $b=0$. Let $G_-$ have $K=\min\mathcal K_n$, member and reference probabilities $r_0=(r_-+r_+)/2$, and nonmember probability $b$. Along $n,m\to\infty$ with $d_b=o(n)$, every deterministic honest procedure has $\E_{G_-}\diam C\ge(1-o(1))D_n$. Consequently,
\[
\mathcal R^*_{\eta,\mathrm{det}}/D_n\longrightarrow1.
\]
At $\eta=\alpha$, $m\asymp n$, $\varepsilon=0$ and fixed rational $b$, the randomised risk has order $\sqrt{n\log n}$.
\end{theorem}
The same response vector can be compatible with many fixed membership assignments. An admissible failure state concentrated on that vector forces a deterministic rule to retain the compatible counts, even at the pure clean law $G_-$. The rational-background condition supplies the compatibility lattice used in this argument. An explicit finite-sample certificate includes $b=0$ and extends to background intervals (Supplement Proposition~\ref{supp-det:finite} and Corollary~\ref{supp-cor:backgroundinterval}).

\begin{theorem}[Randomness from a reference subset]\label{thm:interior}
Fix known $0<\tau<1/2$ with $\tau<r_-<r_+<1-\tau$. There are predetermined reserve sizes $h_m=O_\tau(\log m)$ and fixed linear maps for which the following holds. When $h_m\le m/2$ and $h_m/m\le(r_--\bar b)/2$, reserve a fixed set $S_m$ of $h_m$ references and require
\[
\tau\le p_j(U)\le1-\tau\quad(j\in S_m)\quad\text{in every state}.
\]
At other sizes, including $m=1$, set $S_m=\varnothing$. For this reliability subclass, denoted $\mathcal P_{\eta,\tau,S}$,
\begin{equation}\label{eq:interiorrate}
\mathcal R^*_{\eta,\tau,S,\mathrm{det}}
\asymp\mathcal R^*_{\eta,\tau,S,\mathrm{rand}}
\asymp\text{the right side of \eqref{eq:envrate}}.
\end{equation}
This holds uniformly over the same sample sizes and declarations as Theorem~\ref{thm:environment}, with constants also depending on $\tau$. It also holds when every reference probability lies in $[\tau,1-\tau]$.
\end{theorem}
The interiority condition defines a smaller reliability class in which the reserved responses retain randomness even when calibration fails. The unrestricted class permits deterministic reference responses, which enable the obstruction in Theorem~\ref{thm:deterministic}.

A fixed linear transformation of the reserved responses produces an approximately uniform seed~\cite{zhou2011linear}. Conditional on $U$, the seed is independent of the responses used for inversion. The coverage correction accounts for its approximation error, and the lost reference information is absorbed by the reference-noise term. The matrix certificates and adjusted retained-reference declarations give a finite-sample deterministic procedure on this subclass (Supplement Sections~\ref{supp-sec:reserve_construction}--\ref{supp-sec:reserve_lower}).

\section{Numerical illustrations}\label{sec:chernoff}

\begin{figure}[htbp]\centering
\includegraphics[width=\textwidth,alt={Two panels of normalised finite lower and upper minimax-risk bounds for the Uniform endpoint and Gaussian mean, both approaching the sharp asymptotic limit of one.}]{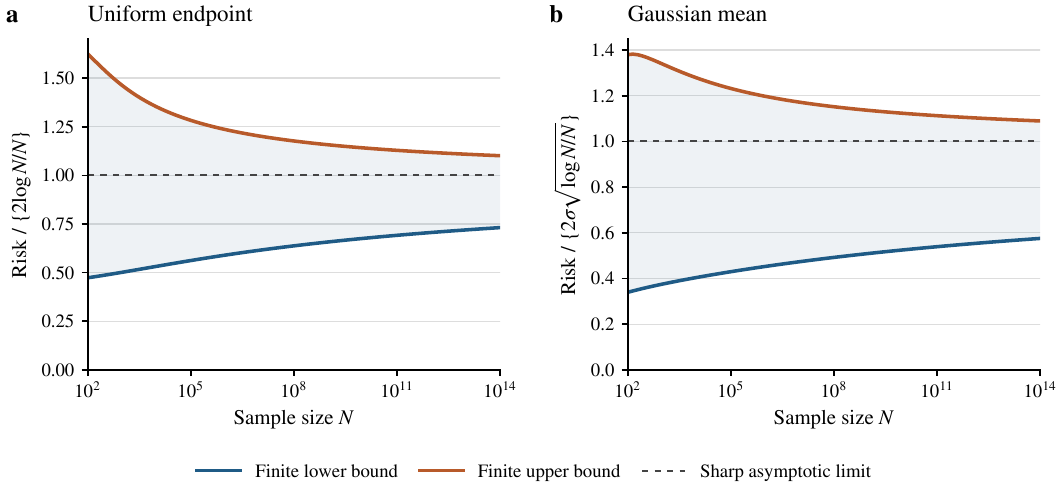}
\caption{Finite-sample bounds for the randomised minimax expected
diameter at the failure boundary. \textbf{a}~Uniform endpoint in $[1,2]$,
normalised by $2\log N/N$. \textbf{b}~Gaussian mean, normalised by
$2\sigma\sqrt{\log N/N}$, with $I=[0,1]$ and $\sigma=1$.
In both panels, $\eta=\alpha=0.05$. Solid curves are finite lower and
upper bounds, shading indicates the interval between them, and the
horizontal dashed line is the sharp asymptotic limit.}
\label{fig:sharp-bounds}
\par\smallskip\noindent{\footnotesize\textbf{Alt text:} Two panels plot normalised finite-sample lower and upper risk bounds against sample size on a logarithmic scale. For both the Uniform endpoint and Gaussian mean, the bounds approach the horizontal asymptotic limit of one from below and above. The gap between the bounds remains visible over the displayed sample-size range.}
\end{figure}

The finite-sample lower and upper bounds bracket the randomised minimax risks and converge slowly to the sharp asymptotic expressions (Figure~\ref{fig:sharp-bounds}). The lower bounds use the exact support comparison and Gaussian testing curve. The upper bounds use the independent full-range correction, with $\beta$ optimised separately at each $N$.

\begin{figure}[htbp]\centering
\includegraphics[width=\textwidth,alt={Finite lower and upper deterministic-risk bounds under polynomial and exponential error-budget slack. Polynomial risk shrinks, while exponential slack yields positive limiting risk.}]{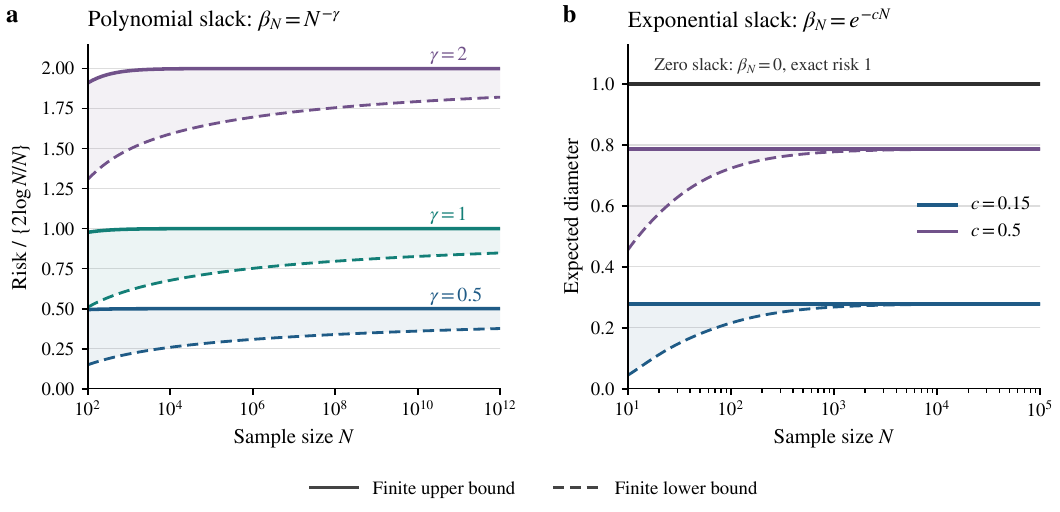}
\caption{Finite-sample deterministic error-budget frontier for the
Uniform endpoint. \textbf{a}~Polynomial effective slack,
$\beta_N=N^{-\gamma}$, with expected diameter normalised by $2\log N/N$.
\textbf{b}~Exponential effective slack, $\beta_N=e^{-cN}$, on the original
diameter scale. The horizontal black line gives the exact risk at zero
slack. Solid and dashed curves are the finite upper and lower bounds,
respectively, with $\beta=(\bar\alpha-\eta)/(1-\eta)$.}
\label{fig:slack-frontier}
\par\smallskip\noindent{\footnotesize\textbf{Alt text:} The left panel gives lower and upper bounds for polynomial effective slack at three values of $\gamma$. The normalised risks approach their respective $\gamma$ values. The right panel gives bounds for two exponential-slack sequences, approaching positive limiting diameters below the horizontal unit-risk line for zero slack.}
\end{figure}

Polynomial effective slack $\beta_N=N^{-\gamma}$ gives deterministic risk $\sim2\gamma\log N/N$, whereas exponential slack $\beta_N=e^{-cN}$ gives a positive limit $\min\{1,2(1-e^{-c})\}$ (Figure~\ref{fig:slack-frontier}). The bands use the finite bounds in Theorem~\ref{slack:sharp} to quantify these regimes at finite sample sizes.

\begin{figure}[H]\centering
\includegraphics[width=\textwidth,alt={Simulated normalised roster confidence-set diameters against failure bound at sample sizes 4000 and 64000, comparing external randomisation, reference reserve and direct inversion.}]{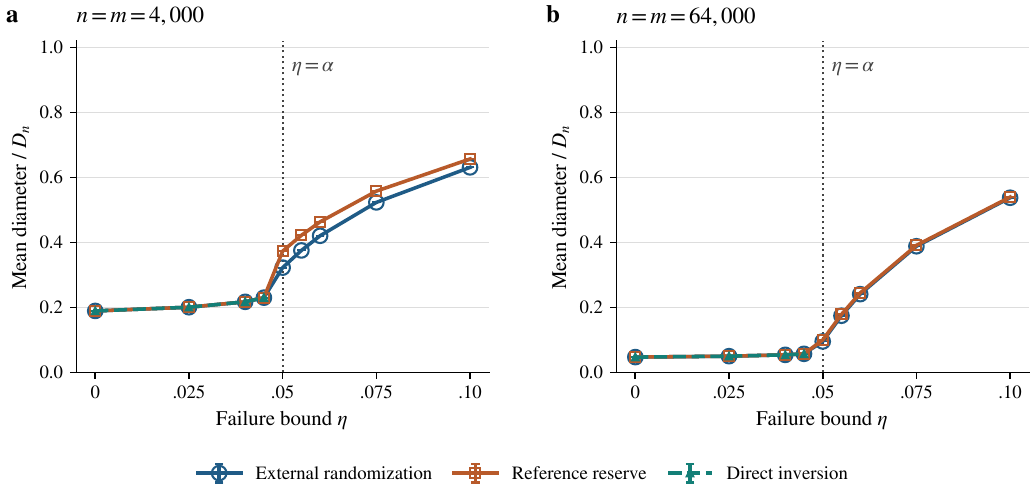}
\caption{Simulated mean roster confidence-set diameter across the declared
failure bound $\eta$. \mbox{\textbf{a} $n=m=4{,}000$}. \mbox{\textbf{b} $n=m=64{,}000$}.
Diameters are normalised by $D_n=0.5n$. Means are under the specified
two-state laws with actual failure probability $\lambda=\eta$.
The dotted vertical line marks $\eta=\alpha=0.05$. External
randomisation and the deterministic reference-reserve procedure use
bound-based selection for $\eta<\alpha$ and the rate choice
$\beta=(s+u)/4$ for $\eta\ge\alpha$. Direct inversion is included only
for $\eta<\alpha$. Error bars are
$\pm1.96$ Monte Carlo standard errors.}
\label{fig:finite}
\par\smallskip\noindent{\footnotesize\textbf{Alt text:} Two panels compare external randomisation, reference reserve and direct inversion at equal main and reference sample sizes of 4,000 and 64,000. The curves coincide below the marked failure bound of 0.05. At and above this bound, the estimated mean diameters increase, with reference-reserve values slightly larger than external-randomisation values. The larger-sample panel has smaller normalised diameters.}
\end{figure}

\Needspace{6\baselineskip}
For roster computation, the Chernoff/H refinement of the Hoeffding inversion $C_H$ gives a contained set with the same clean error allocation. The external and internal corrections then preserve the coverage bound (Supplement Section~\ref{supp-sec:chernoff}). The generating mixtures in Figure~\ref{fig:finite} have $\alpha=0.05$, $\tau=0.1$, known background $b=0.1$, $[r_-,r_+]=[0.45,0.55]$, $\varepsilon=0$, $\mathcal K_n=\{0.2n,\ldots,0.7n\}$ and true count $K=0.4n$. Their actual failure probability is set to the declared bound, $\lambda=\eta$. Each plotted mean estimates the expected diameter under that specified mixture law.

For $\eta<\alpha$, each procedure selects $\beta$ before observing the data by minimising a finite-sample diameter bound over a fixed grid, including direct inversion with all references as a candidate. For $\eta\ge\alpha$, the procedures use the rate construction $\beta=(s+u)/4$. The curves across $\eta=\alpha$ therefore reflect both the change in failure bound and this change in selection rule.

At $\eta=\alpha$, external randomisation and the reference-reserve rule have estimated mean diameters $0.3225D_n$ and $0.3732D_n$ at $n=m=4{,}000$, decreasing to $0.0955D_n$ and $0.0999D_n$ at $64{,}000$. The respective reference reserves contain 126 and 168 responses. At both sizes in Figure~\ref{fig:finite}, the bound-based selection chooses direct inversion at every displayed $\eta<\alpha$, so the three curves coincide there.

At the pure clean law $G_-$ with $n=m=4{,}000$, the finite-sample lower bound on expected diameter for every deterministic procedure honest over the unrestricted reliability class is approximately $0.990D_n$. At the same generating law, external randomisation has estimated mean diameter $0.217D_n$, and the deterministic rule honest over the reference-interior subclass has estimated mean diameter $0.235D_n$ (Supplement Tables~\ref{supp-tab:samelaw}--\ref{supp-tab:diameter_distribution}).

\begin{samepage}
\section{Discussion}

At the failure boundary, honest precision depends on the information in the clean experiment and the randomness preserved by the failure class. An independent seed remains valid under arbitrary joint failures. Observation order supplies such randomness under exchangeability without ties, and a reference reserve supplies an approximate seed under conditional independence and interiority. These restrictions explain how a deterministic function of the data can attain randomised precision on a smaller failure class. The choice of failure class therefore specifies both the departures allowed from the clean model and the random sources available for inference.
\par\end{samepage}

Learning the reliability and calibration declarations from auxiliary data raises a further error-budget problem. A confidence bound on the probability of calibration failure introduces uncertainty into the declared mixture class. How should the total noncoverage budget be allocated between validating that declaration and constructing the final confidence set? Near the failure boundary, the optimal allocation may depend on the auxiliary sample size as well as the information in the main experiment. A related question is whether aggregate conditions on reserved responses can preserve a uniformly accurate seed while allowing some conditional response probabilities to approach zero or one. Such conditions would connect the amount of randomness retained in failure states to the reference information needed for calibration.

\section*{Acknowledgement}
We thank Professors Yang Li (Renmin University of China), Ye Shen (University of Georgia), David Hongxiang Qiu (Michigan State University), and Zhanrui Cai (The University of Hong Kong) for their insightful comments and constructive suggestions.

\label{main:body-end}

\clearpage

\clearpage
\setcounter{section}{0}
\setcounter{theorem}{0}
\setcounter{proposition}{0}
\setcounter{corollary}{0}
\setcounter{lemma}{0}
\setcounter{remark}{0}
\setcounter{table}{0}
\setcounter{figure}{0}
\setcounter{equation}{0}
\renewcommand{\thesection}{S\arabic{section}}
\renewcommand{\thetheorem}{S\arabic{theorem}}
\renewcommand{\theproposition}{S\arabic{proposition}}
\renewcommand{\thecorollary}{S\arabic{corollary}}
\renewcommand{\thelemma}{S\arabic{lemma}}
\renewcommand{\theremark}{S\arabic{remark}}
\renewcommand{\thetable}{S\arabic{table}}
\renewcommand{\thefigure}{S\arabic{figure}}
\renewcommand{\theequation}{S\arabic{equation}}
% Hyperref destination names must also be unique after the counters reset.
\renewcommand{\theHsection}{supp.\arabic{section}}
\renewcommand{\theHsubsection}{supp.\arabic{section}.\arabic{subsection}}
\renewcommand{\theHequation}{supp.\arabic{equation}}
\renewcommand{\theHtable}{supp.\arabic{table}}
\renewcommand{\theHfigure}{supp.\arabic{figure}}
\providecommand{\theHtheorem}{}
\renewcommand{\theHtheorem}{supp.\arabic{theorem}}
\providecommand{\theHlemma}{}
\renewcommand{\theHlemma}{supp.\arabic{lemma}}
\providecommand{\theHproposition}{}
\renewcommand{\theHproposition}{supp.\arabic{proposition}}
\providecommand{\theHcorollary}{}
\renewcommand{\theHcorollary}{supp.\arabic{corollary}}
\begin{center}
{\Large\bfseries Honest confidence sets under experiment-level failure}\\[0.6em]
{\large Supplementary Material}
\end{center}
\smallskip
This supplement gives the complete proofs, finite-sample bounds and numerical methods.
Local results, equations and sections carry the prefix S, and references to the main text are identified explicitly.
\par\medskip\noindent\textbf{Supplementary contents}\par\smallskip
\begingroup\small
\noindent S1\quad Proofs for the Uniform endpoint experiment\dotfill\pageref{supp-sec:endpoint-proofs}\par
\noindent S2\quad Proof of the sharp Gaussian theorem\dotfill\pageref{supp-gaussian:proofs}\par
\noindent S3\quad Proof of the deterministic error-budget frontier\dotfill\pageref{supp-slack:proofs}\par
\noindent S4\quad The coverage budget and finite information bound\dotfill\pageref{supp-sec:general-finite}\par
\noindent S5\quad Proof of the critical information rate\dotfill\pageref{supp-sec:gp-rate-proof}\par
\noindent S6\quad Roster conventions\dotfill\pageref{supp-sec:roster-conventions}\par
\noindent S7\quad Statewise roster coverage and the baseline rate\dotfill\pageref{supp-sec:roster-baseline}\par
\noindent S8\quad Roster rate and randomisation proofs\dotfill\pageref{supp-sec:roster-proofs}\par
\noindent S9\quad Computation\dotfill\pageref{supp-sec:chernoff}\par
\noindent S10\quad Numerical results\dotfill\pageref{supp-sec:numerical-results}\par
\endgroup
\clearpage
\section{Proofs for the Uniform endpoint experiment}\label{supp-sec:endpoint-proofs}

We use the experiment and failure classes of main Section~\ref{ep:section}.
\begin{proposition}[Finite endpoint bounds]\label{supp-ep:finitebounds}
Put $c_\alpha=\alpha/(1-\alpha)$. For $1\le a<2$, $0<h<2-a$, and $L=2-a-h$, every honest procedure over each of the three failure classes satisfies
\begin{equation}\label{supp-ep:finitelower}
R_N(C;\mathcal Q_N)\ge\frac{hL}{L+c_\alpha h\{1+\exp(Nh/a)\}}.
\end{equation}
The construction in main Theorem~\ref{ep:sharp} has risk at most
\begin{equation}\label{supp-ep:finiteupper-copy}
p_\beta+(1-p_\beta)\min\{1,2(1-\beta^{1/N})\}.
\end{equation}
In particular, the risk is at least $\log N/(4N)$ for $N\ge2$.
\end{proposition}

\begin{proof}[Proof of main Theorem~\ref{ep:sharp} and Proposition~\ref{supp-ep:finitebounds}]
\textit{Upper bound.} Put $t_\beta=\beta^{1/N}$. Under $G_{\theta,N}$, the event $\theta\in B_\beta(M_N)$ is $M_N\ge\theta t_\beta$, so its probability is exactly $1-\beta$. If $B_\beta$ is nonempty with endpoints $L_B,U_B$, then $L_B\ge M_N\ge t_\beta U_B$ and $U_B\le2$. Hence
\[
\diam B_\beta\le(1-t_\beta)U_B\le2(1-t_\beta).
\]
The interval is contained in $[1,2]$, giving $\diam B_\beta\le D_{N,\beta}$ for every $x\in[0,2]^N$, including empty outputs.

Under~\eqref{ep:model}, the base error is at most
\[
(1-\lambda)\beta+\lambda
\le q_\beta:=\alpha+(1-\alpha)\beta.
\]
The independent full-range decision reduces this bound to $(1-p_\beta)q_\beta=\alpha$. Its expected diameter is at most~\eqref{ep:finiteupper}. For fixed $\alpha$ and $\beta=1/N$,
\[
p_\beta=O_\alpha(N^{-1}),\qquad
2(1-N^{-1/N})=\frac{2\log N}{N}
+O\!\left(\frac{(\log N)^2}{N^2}\right).
\]
This proves the upper asymptotic bound over all three classes.

\par\smallskip\noindent\textit{A shared failure law.} Fix an honest procedure and abbreviate its risk by $R$. Write
\[
e_\theta=\Prb_{G_{\theta,N},\xi}\{\theta\notin C\}.
\]
The law $G_{1,N}$ belongs to each failure class. Honesty at $\lambda=\alpha$ and $Q=G_{1,N}$ gives
\[
(1-\alpha)e_\theta
\le\alpha\Prb_{G_{1,N},\xi}\{\theta\in C\}.
\]
By Tonelli's theorem and $\operatorname{Leb}_1(C)\le\diam C$,
\begin{equation}\label{supp-ep:budget}
\int_1^2e_\theta\,d\theta
\le c_\alpha\E_{G_{1,N},\xi}\operatorname{Leb}_1(C)
\le c_\alpha R.
\end{equation}
Here $G_{1,N}$ is also an allowed clean law, at $\theta=1$ and $\lambda=0$, so its expected diameter is bounded by $R$ without a mixture-weight factor.

\par\smallskip\noindent\textit{Nested supports.} For $1\le\theta\le2-h$, the densities give the measure inequality
\begin{equation}\label{supp-ep:domination}
G_{\theta+h,N}\ge
\left(\frac{\theta}{\theta+h}\right)^NG_{\theta,N}.
\end{equation}
The same inequality holds after adjoining the independent seed. For $\theta\ge a$ it yields
\[
\Prb_{G_{\theta,N},\xi}\{\theta+h\notin C\}
\le(1+h/\theta)^Ne_{\theta+h}
\le e^{Nh/a}e_{\theta+h}.
\]
A set containing both $\theta$ and $\theta+h$ has diameter at least $h$. Therefore
\[
R\ge h\{1-e_\theta-e^{Nh/a}e_{\theta+h}\}.
\]
Integrating over $[a,2-h]$ and applying~\eqref{supp-ep:budget} gives
\[
LR\ge h\{L-c_\alpha R(1+e^{Nh/a})\},
\]
which proves~\eqref{supp-ep:finitelower}.

For the sharp lower constant, we localise near the upper endpoint. The subtraction of $2\log s_N$ below makes the exponential contribution to the denominator negligible relative to $L_N$. Put $s_N=\log N$ and, for all sufficiently large $N$, choose
\[
a_N=2-s_N^{-1/2},\qquad
h_N=\frac{a_N(s_N-2\log s_N)}N,\qquad
L_N=2-a_N-h_N.
\]
Then $L_N\sim s_N^{-1/2}$, $e^{Nh_N/a_N}=N/s_N^2$, and
\[
\frac{c_\alpha h_N(1+e^{Nh_N/a_N})}{L_N}
=O_\alpha(s_N^{-1/2})\longrightarrow0.
\]
Equation~\eqref{supp-ep:finitelower} gives $R\ge(1-o(1))h_N$, while $Nh_N/\log N\to2$. These choices do not depend on the procedure, so the bound also holds after taking the infimum. This proves~\eqref{ep:sharpconstant}.

For a fixed parameter interval $[\ell,u]\subset(0,\infty)$ with $D=u-\ell>0$, use $G_{\ell,N}$ as the shared failure law and replace $L$ in~\eqref{supp-ep:finitelower} by $u-a-h$. Taking $a_N=u-D/\sqrt{s_N}$ and the same $h_N$ makes the denominator correction $o(L_N)$ and gives $Nh_N/\log N\to u$. The clipped construction has risk at most $p_\beta D+(1-p_\beta)\min\{D,u(1-\beta^{1/N})\}$, which matches this lower bound at $\beta=1/N$.

\par\smallskip\noindent\textit{Deterministic procedures.} Let $C=C(X)$ be honest over $\mathcal Q_N^{\mathrm{ac}}$. If
\[
E_\theta=\{x:\theta\notin C(x)\},\qquad
e_\theta=G_{\theta,N}(E_\theta)>0,
\]
choose the failure law $Q_\theta=G_{\theta,N}(\,\cdot\mid E_\theta)$. The minimax supremum permits this choice after fixing the deterministic rule. It is absolutely continuous, with density
\[
q_\theta(x)=\frac{\theta^{-N}}{e_\theta}
\mathbf1\{x\in[0,\theta]^N\}\mathbf1\{x\in E_\theta\}.
\]
Its error probability is one. At $\lambda=\alpha$, the mixture error is therefore $(1-\alpha)e_\theta+\alpha>\alpha$, a contradiction. Hence $e_\theta=0$ for each fixed $\theta$.

In particular, $G_{1,N}\{1\in C\}=G_{2,N}\{2\in C\}=1$. Since $G_{1,N}\ll G_{2,N}$, both endpoints belong to $C$ with $G_{1,N}$-probability one. Hence $\E_{G_{1,N}}\diam C=1$. The full-range rule attains one, proving~\eqref{ep:det} for the absolutely continuous and unrestricted classes.
\end{proof}
\begin{proof}[Proof of main Proposition~\ref{ep:exchangeable}]
Put $r_{N,\beta}=\lceil Np_\beta\rceil$ and $f_{N,\beta}=r_{N,\beta}/N$. Under any exchangeable law with no ties, permutation symmetry gives, for every Borel set $B$,
\[
\Prb\{J=j,M_N\in B\}=N^{-1}\Prb\{M_N\in B\},
\qquad j=1,\ldots,N.
\]
Indeed, permutations preserve $M_N$ and give equal probabilities to the $N$ disjoint possible maximum indices. Thus $J$ is uniform and independent of $M_N$.

Both $G_{\theta,N}$ and every $Q\in\mathcal Q_N^{\mathrm{ex}}$ have this property. If $S$ denotes the mixture state in~\eqref{ep:model}, then
\[
\Prb\{J=j,M_N\in B,S=s\}
=N^{-1}\Prb\{M_N\in B,S=s\}.
\]
The common uniform law of $J$ therefore makes it independent of $(M_N,S)$, and in particular of $M_N$ after mixing. This justifies the factorisation used by the deterministic rule under every allowed mixture.

Since $f_{N,\beta}\ge p_\beta$, its noncoverage probability is at most
\[
(1-f_{N,\beta})\{\lambda+(1-\lambda)\beta\}
\le(1-p_\beta)q_\beta=\alpha.
\]
Independence of $J$ and $M_N$, the pointwise diameter bound, and $0\le f_{N,\beta}-p_\beta\le N^{-1}$ give~\eqref{ep:internalfinite}. Take $\beta=1/N$ for the upper constant two. The lower bound in Theorem~\ref{ep:sharp} applies because $G_{1,N}\in\mathcal Q_N^{\mathrm{ex}}$, and deterministic rules are a subclass of randomised rules.
\end{proof}
\label{supp-sec:endpoint-additional}

\subsection{An explicit finite-sample lower bound}

For $N\ge2$, take $a=1$ and $h=\log N/(2N)$ in~\eqref{supp-ep:finitelower}. These choices give $0<h<1$ and $L=1-h$. Since $c_\alpha=\alpha/(1-\alpha)<1$ and
\[
2h+h\sqrt N
=\frac{\log N}{N}+\frac{\log N}{2\sqrt N}
\le\frac2{\mathrm e}<1,
\]
we have $c_\alpha h(1+\sqrt N)\le1-h=L$. The two terms above are at most $1/\mathrm e$ each, by maximising $\log x/x$ and $\log x/(2\sqrt x)$ over $x\ge1$. Also $\exp(Nh/a)=\sqrt N$, so~\eqref{supp-ep:finitelower} gives
\[
R_N(C;\mathcal Q_N)
\ge\frac{hL}{L+c_\alpha h(1+\sqrt N)}
\ge\frac h2=\frac{\log N}{4N}.
\]

\subsection{Permutation-invariant deterministic rules}

\begin{proof}[Proof of Corollary~\ref{ep:invariant}]
Fix such an honest rule. For each $\theta$, its error event $E_\theta=\{x:\theta\notin C(x)\}$ is invariant under coordinate permutations. If $G_{\theta,N}(E_\theta)>0$, the conditional law $Q_\theta=G_{\theta,N}(\,\cdot\mid E_\theta)$ remains exchangeable and has no ties. Thus $Q_\theta\in\mathcal Q_N^{\mathrm{ex}}$, and the critical mixture has error $\alpha+(1-\alpha)G_{\theta,N}(E_\theta)>\alpha$, contradicting honesty. Consequently the clean error is zero at both $\theta=1$ and $\theta=2$. The relation $G_{1,N}\ll G_{2,N}$ then gives $\{1,2\}\subseteq C$ with $G_{1,N}$-probability one, so the risk is one. The full-range rule is invariant and attains this value.
\end{proof}

\section{Proof of the sharp Gaussian theorem}\label{supp-gaussian:proofs}

This section proves Theorem~\ref{gaussian:sharp} and gives finite-sample bounds. Fix $I=[a,b]$, $D=b-a>0$, $\sigma>0$ and $0<\alpha<1/2$. All three failure classes are as defined in the main text, on $\mathbb R^N$. Sets have a jointly measurable inclusion graph and measurable diameter, with $\diam\varnothing=0$. An auxiliary seed has the same law throughout the model and is independent of observations and the failure state. For an honest rule $C$, write $R$ for its worst-case expected diameter over the allowed mixture laws and set
\[
e_t=\Prb_{G_{t,N},\xi}\{t\notin C\},\qquad
c_\alpha=\frac{\alpha}{1-\alpha}.
\]

\subsection{Finite lower bound and the sharp constant}

For every $0<d<D/2$ and $0<\varepsilon<1$,
\begin{equation}\label{supp-gaussian:finitelower}
R\ge\min\left\{
\frac{\varepsilon(D-2d)}{2c_\alpha},\quad
2d\left[1-2\Phi\left\{\Phi^{-1}(\varepsilon)
              +\frac{\sqrt N d}{\sigma}\right\}\right]_+
\right\}.
\end{equation}
To prove this, use the common failure law $G_{a,N}$, which belongs to all three classes. Honesty at $\lambda=\alpha$ gives
\[
(1-\alpha)e_t
\le\alpha\Prb_{G_{a,N},\xi}\{t\in C\}.
\]
Tonelli's theorem and $\operatorname{Leb}(C)\le\diam C$ imply
\begin{equation}\label{supp-gaussian:budget}
\int_a^b e_t\,dt\le
c_\alpha\E_{G_{a,N},\xi}\operatorname{Leb}(C)
\le c_\alpha R.
\end{equation}
The last inequality uses $G_{a,N}$ itself as an admissible clean law.

Let $A_\varepsilon=\{t\in I:e_t>\varepsilon\}$. Equation~\eqref{supp-gaussian:budget} bounds its Lebesgue measure by $c_\alpha R/\varepsilon$. Among the possible centres $\theta\in[a+d,b-d]$, the centres for which $\theta-d\in A_\varepsilon$ or $\theta+d\in A_\varepsilon$ have total measure at most $2c_\alpha R/\varepsilon$. Thus, if
\begin{equation}\label{supp-gaussian:goodcentercondition}
R<\frac{\varepsilon(D-2d)}{2c_\alpha},
\end{equation}
there exists a centre $\theta$ with $e_{\theta-d}\le\varepsilon$ and $e_{\theta+d}\le\varepsilon$.

The Gaussian testing curve transfers these two error bounds to the centre. For any event $E$ in the observation and seed space,
\begin{equation}\label{supp-gaussian:testing}
\Prb_{G_{t,N},\xi}(E)\le\varepsilon
\quad\Longrightarrow\quad
\Prb_{G_{\theta,N},\xi}(E)
\le\Phi\left\{\Phi^{-1}(\varepsilon)
                  +\frac{\sqrt N|t-\theta|}{\sigma}\right\}.
\end{equation}
Indeed, for $t>\theta$, the likelihood ratio $dG_{\theta,N}/dG_{t,N}$ decreases with $\bar X$. Among events of $G_{t,N}$-probability at most $\varepsilon$, the Neyman--Pearson lemma maximises the $G_{\theta,N}$-probability at
\[
\left\{\bar X\le t+\frac{\sigma}{\sqrt N}
                       \Phi^{-1}(\varepsilon)\right\}.
\]
Its $G_{\theta,N}$-probability is the right side of \eqref{supp-gaussian:testing}. For $t<\theta$, the maximising event is an upper tail and gives the same formula. The case $t=\theta$ is immediate. Adjoining the common independent seed leaves the likelihood ratio unchanged, so the argument includes randomised events.

Apply~\eqref{supp-gaussian:testing} to the two exclusion events for $t=\theta-d$ and $t=\theta+d$. Under $G_{\theta,N}$, both points belong to $C$ with probability at least
\[
\left[1-2\Phi\left\{\Phi^{-1}(\varepsilon)
                     +\frac{\sqrt N d}{\sigma}\right\}\right]_+.
\]
On this event $\diam C\ge2d$, proving the second term of \eqref{supp-gaussian:finitelower} when \eqref{supp-gaussian:goodcentercondition} holds. If it fails, $R$ already exceeds the first term. This proves the finite lower bound for every honest rule and hence for the minimax risk.

For the asymptotic constant, choose $\varepsilon_N$ so that the first alternative in~\eqref{supp-gaussian:finitelower} exceeds the target rate, while $d_N$ makes both transferred testing errors vanish. Fix $0<\delta<1$ and, for sufficiently large $N$, take
\[
\varepsilon_N=N^{-1/2}\log N,\qquad
d_N=(1-\delta)\sigma\sqrt{\frac{\log N}{N}}.
\]
These choices satisfy the finite-bound conditions. The first term in \eqref{supp-gaussian:finitelower}, divided by $\sqrt{\log N/N}$, tends to infinity. Gaussian quantile asymptotics give
\[
\frac{\Phi^{-1}(\varepsilon_N)}{\sqrt{\log N}}\longrightarrow-1,
\qquad
\Phi^{-1}(\varepsilon_N)+\frac{\sqrt N d_N}{\sigma}
=(-\delta+o(1))\sqrt{\log N}\longrightarrow-\infty.
\]
Consequently the second term is $(2(1-\delta)\sigma+o(1))\sqrt{\log N/N}$. The choices are independent of the rule, so taking the infimum and then letting $\delta\downarrow0$ yields the sharp lower constant $2\sigma$.

\subsection{An upper bound valid under every failure law}

For $0<\beta<1$, set
\[
z_\beta=\Phi^{-1}(1-\beta/2),\qquad
w_{N,\beta}=\min\left\{D,\frac{2\sigma z_\beta}{\sqrt N}\right\},
\qquad
p_\beta=\frac{(1-\alpha)\beta}{\alpha+(1-\alpha)\beta}.
\]
The clipped interval $B_\beta$ in~\eqref{gaussian:base} has exact clean noncoverage $\beta$ because $\theta\in I$ and the event $\theta\in B_\beta$ is precisely $|\bar X-\theta|\le\sigma z_\beta/\sqrt N$. At every observation vector, its diameter is at most $w_{N,\beta}$. Return $I$ with independent probability $p_\beta$, and $B_\beta$ otherwise. The noncoverage probability is at most
\[
(1-p_\beta)\{(1-\lambda)\beta+\lambda\}
\le(1-p_\beta)\{(1-\alpha)\beta+\alpha\}=\alpha.
\]
Its risk over the allowed mixture laws therefore satisfies
\begin{equation}\label{supp-gaussian:finiteupper}
R\le p_\beta D+(1-p_\beta)w_{N,\beta}.
\end{equation}
For $N\ge2$, take $\beta=N^{-1/2}$. Then $p_\beta=O(N^{-1/2})$ and $z_\beta/\sqrt{\log N}\to1$. Combining~\eqref{supp-gaussian:finiteupper} with the preceding lower bound proves~\eqref{gaussian:sharpconstant}.

\subsection{Deterministic rules and exchangeability}

First consider a deterministic rule honest under absolutely continuous failures. If the clean error event $E_\theta=\{x:\theta\notin C(x)\}$ has probability $e_\theta>0$, the law
\[
Q_\theta=G_{\theta,N}(\,\cdot\mid E_\theta)
\]
is absolutely continuous and assigns probability one to that event. At $\lambda=\alpha$, noncoverage would be $(1-\alpha)e_\theta+\alpha>\alpha$. Thus $e_\theta=0$ for every $\theta\in I$. In particular, $a\in C$ almost surely under $G_{a,N}$ and $b\in C$ almost surely under $G_{b,N}$. The two Gaussian laws are mutually absolutely continuous, so both statements hold almost surely under $G_{a,N}$. The diameter equals $D$ there. Returning $I$ attains risk $D$, which proves the deterministic claim for the absolutely continuous and unrestricted classes.

For exchangeable failures without ties, define $J=\arg\max_i X_i$, with an arbitrary fixed convention on ties. For every Borel set $H\subseteq\mathbb R$, symmetry gives
\begin{equation}\label{supp-gaussian:indexindependence}
\Prb\{J=j,\bar X\in H\}
=\frac1N\Prb\{\bar X\in H\},\qquad j=1,\ldots,N.
\end{equation}
To see this, coordinate permutations preserve $\bar X$ and permute the $N$ disjoint maximum-index events transitively. Their probabilities are equal and sum to $\Prb\{\bar X\in H\}$. Thus $J$ is uniform and independent of $\bar X$ under each clean and failure law. If $S$ records the mixture state, the same identity with $S=s$ adjoined shows that $J$ is independent of $(\bar X,S)$.

Put $k_{N,\beta}=\lceil Np_\beta\rceil$ and $f_{N,\beta}=k_{N,\beta}/N$. The deterministic rule returns $I$ if $J\le k_{N,\beta}$ and $B_\beta(\bar X)$ otherwise. Equation~\eqref{supp-gaussian:indexindependence} bounds its noncoverage by
\[
(1-f_{N,\beta})\{(1-\lambda)\beta+\lambda\}
\le(1-p_\beta)\{(1-\alpha)\beta+\alpha\}=\alpha.
\]
Its risk over the allowed mixture laws is at most
\begin{equation}\label{supp-gaussian:internalfinite}
f_{N,\beta}D+(1-f_{N,\beta})w_{N,\beta}
\le p_\beta D+(1-p_\beta)w_{N,\beta}+\frac DN.
\end{equation}
Taking $\beta=N^{-1/2}$ proves the deterministic sharp constant under exchangeability, since the lower bound for randomised rules also applies to deterministic rules.

Finally, if $C$ is permutation invariant, each $E_\theta$ is an invariant event. Whenever it has positive clean probability, $G_{\theta,N}(\,\cdot\mid E_\theta)$ remains exchangeable and has no ties. It is therefore an admissible failure law. The preceding zero-error argument applies and gives diameter $D$ almost surely under $G_{a,N}$. The full-range rule is itself invariant and attains $D$, completing the proof of Theorem~\ref{gaussian:sharp}.

\section{Proof of the deterministic error-budget frontier}\label{supp-slack:proofs}

We prove Theorem~\ref{slack:sharp} under the three settings stated there. Fix a deterministic rule and write
\[
E_\theta=\{x:\theta\notin C(x)\},\qquad
e_\theta=G_{\theta,N}(E_\theta),\qquad \theta\in[1,2].
\]
Whenever $e_\theta>0$, the conditioned clean law $Q_\theta=G_{\theta,N}(\,\cdot\mid E_\theta)$ is absolutely continuous. If the rule is permutation invariant, it is also exchangeable and has no ties. Thus it belongs to the relevant failure class, and honesty at $\lambda=\eta$ implies
\begin{equation}\label{supp-slack:conditionalerror}
\eta+(1-\eta)e_\theta\le\bar\alpha
\qquad\text{whenever }e_\theta>0.
\end{equation}

\subsection{Exhausted budget and the exact clean-error requirement}

If $\eta\ge\bar\alpha$,~\eqref{supp-slack:conditionalerror} forces $e_\theta=0$ for every $\theta$. In particular, $1\in C$ almost surely under $G_{1,N}$ and $2\in C$ almost surely under $G_{2,N}$. Since $G_{1,N}\ll G_{2,N}$, both endpoints belong to $C$ almost surely under $G_{1,N}$, giving risk at least one. The full-range rule attains one, proving~\eqref{slack:exhausted}.

Suppose now that $\eta<\bar\alpha$ and let $\beta=(\bar\alpha-\eta)/(1-\eta)$. Equation~\eqref{supp-slack:conditionalerror} implies $e_\theta\le\beta$ for every $\theta$, including parameters with $e_\theta=0$. Conversely, any deterministic rule with these clean-error bounds satisfies, for every admissible failure law,
\[
\Prb_{P_{\theta,\lambda,Q}}\{\theta\notin C\}
\le(1-\lambda)\beta+\lambda
\le(1-\eta)\beta+\eta=\bar\alpha.
\]
Thus robust honesty is equivalent to clean noncoverage at most $\beta$ in the stated classes of deterministic rules. This also covers $\eta=0$, when the equivalence follows directly from the clean model.

\subsection{Matching diameter bounds}

Set $t=\log(1/\beta)$ and $D_N(t)=\min\{1,2(1-e^{-t/N})\}$. With $M_N=\max_iX_i$, the base interval is
\[
B_\beta(M_N)=
[\max\{1,M_N\},\min\{2,M_N/\beta^{1/N}\}],
\]
with reversed intervals interpreted as empty. Under $G_{\theta,N}$, it contains $\theta$ precisely when $M_N\ge\theta\beta^{1/N}$ and hence has clean error $\beta$. For any nonempty output, its endpoints $L_B,U_B$ satisfy $L_B\ge M_N\ge\beta^{1/N}U_B$ and $U_B\le2$. It follows that
\[
\diam B_\beta\le
\min\{1,2(1-\beta^{1/N})\}=D_N(t)
\]
at every observation vector. The rule is permutation invariant, so it proves the upper bound in all three settings.

For the lower bound, let $0<s<t$ and choose
\[
u=\max\{1,2e^{-(t-s)/N}\}\in[1,2).
\]
The nested uniform supports give the measure inequality $G_{2,N}\ge(u/2)^N G_{u,N}$. Since the clean error at $2$ is at most $\beta=e^{-t}$,
\[
G_{u,N}\{2\notin C\}
\le(2/u)^N G_{2,N}\{2\notin C\}
\le(2/u)^Ne^{-t}\le e^{-s}.
\]
The last inequality is an equality when $u>1$. When $u=1$, its definition gives $t-s\ge N\log2$, which yields the same inequality. Also $G_{u,N}\{u\notin C\}\le e^{-t}$. Under $G_{u,N}$, therefore, both $u$ and $2$ belong to $C$ with probability at least $[1-e^{-t}-e^{-s}]_+$. Since $2-u=D_N(t-s)$, every admissible rule satisfies
\[
R_N(C)\ge\E_{G_{u,N}}\diam C
\ge D_N(t-s)[1-e^{-t}-e^{-s}]_+.
\]
Taking the infimum proves~\eqref{slack:finite}.

Finally, $D_N$ is nondecreasing and concave on $[0,\infty)$, with $D_N(0)=0$, because its derivative decreases until $N\log2$ and drops to zero there. For $t>1$, put $s=\sqrt t$. Concavity gives
\[
\frac{D_N(t-\sqrt t)}{D_N(t)}\ge1-t^{-1/2}.
\]
For all sufficiently large $t$, the finite bounds imply
\begin{equation}\label{supp-slack:relativebound}
1\ge\frac{\mathcal R_N^{\mathrm{det}}(\eta,\bar\alpha;\mathcal Q)}{D_N(t)}
\ge (1-t^{-1/2})(1-e^{-t}-e^{-\sqrt t}).
\end{equation}
The right side tends to one as $t\to\infty$, independently of $N$. This proves~\eqref{slack:asymptotic}. Substituting $\beta_N=\delta_N/(1-\alpha)$ gives the three coverage-slack regimes in~\eqref{slack:regimes}. When the allowed error stays at $\alpha$ and $\eta_N=\alpha-\delta_N$, the effective budget becomes $\beta_N=\delta_N/(1-\alpha+\delta_N)$ and gives the same leading limits.

\section{The coverage budget and finite information bound}\label{supp-sec:general-finite}

Write $\nu$ for the common law of the independent seed $\xi$.
\begin{proof}[Proof of main Lemma~\ref{lem:gp-budget}]

Coverage under the critical mixture gives, for each $\theta$,
\begin{equation}
\label{supp-eq:gp-pointwise-budget}
  (1-\alpha)e_\theta
  \le \alpha\Prb_{Q_N\otimes\nu}\{\theta\in C\}.
\end{equation}
For any admissible parameter, the risk of its critical mixture is at least $\alpha\E_{Q_N\otimes\nu}\diam C$.  Thus $\alpha\E_{Q_N\otimes\nu}\diam C\le R$. Tonelli's theorem and $\operatorname{Leb}(C)\le\diam C$ prove \eqref{eq:gp-budget-continuous}.  For a grid, the pointwise inequality
\[
  \Delta_N|C\cap\Theta_N|\le\diam C+\Delta_N
\]
proves \eqref{eq:gp-budget-grid}.  When $Q_N$ itself is an allowed observation law, the bound $\E_{Q_N\otimes\nu}\diam C\le R$ gives the two stronger inequalities by substitution.
\end{proof}
For the next result, set $\Delta_N=0$ in the interval case.  Let $b_\alpha=1/(1-\alpha)$, or $b_\alpha=c_\alpha$ when the stronger condition in Lemma~\ref{lem:gp-budget} holds.  The budget is then $b_\alpha R+c_\alpha\Delta_N$ in both cases.

\begin{theorem}[Finite-sample lower bound via one-sided KL control]
\label{supp-thm:gp-finite}
Fix $d>0$ and a target set $T_N\subseteq\Theta_N$.  In the interval case, let $T_N$ be measurable with length $\ell_N>0$. In the grid case, put $\ell_N=\Delta_N|T_N|>0$. Suppose that, for every $t\in T_N$, there is a clean parameter $s\in\Theta_N$ such that
\begin{equation}
\label{supp-eq:gp-directed-neighbor}
  |s-t|\ge d,
  \qquad
  \KL(G_{s,N}\Vert G_{t,N})\le K_N(d)<\infty.
\end{equation}
Every honest $C$ with $R<(1-\alpha)d$ satisfies
\begin{equation}
\label{supp-eq:gp-finite-budget}
  b_\alpha R+c_\alpha\Delta_N
  \ge
  \ell_N\exp\!\left\{
       -\frac{K_N(d)+\log2}{1-\alpha-R/d}
               \right\}.
\end{equation}
In particular, with $a_\alpha=(1-\alpha)/2$ and $[z]_+=\max(z,0)$,
\begin{equation}
\label{supp-eq:gp-finite-lower}
 R_N(C)\ge
 \min\!\left\{
   a_\alpha d,
   \frac{\bigl[\ell_N
       e^{-(K_N(d)+\log2)/a_\alpha}
       -c_\alpha\Delta_N\bigr]_+}{b_\alpha}
          \right\}.
\end{equation}
\end{theorem}

\begin{proof}
Fix $t\in T_N$ and a partner $s$ from \eqref{supp-eq:gp-directed-neighbor}.  Under $G_{s,N}\otimes\nu$, the event $E_t=\{t\notin C\}$ has probability at least
\[
  \Prb\{s\in C,\ \diam C<d\}
  \ge 1-e_s-R/d
  \ge 1-\alpha-R/d=:a>0.
\]
Here clean honesty gives $e_s\le\alpha$, and Markov's inequality uses $\E_{G_{s,N}\otimes\nu}\diam C\le R$. Under $G_{t,N}\otimes\nu$, the same event has probability $e_t$. Adding the common independent seed leaves the KL divergence unchanged. If $p=\Prb_{G_{s,N}\otimes\nu}(E_t)$, binary data processing gives
\[
  K_N(d)\ge \operatorname{kl}(p,e_t)
     \ge p\log(1/e_t)-\log2
     \ge a\log(1/e_t)-\log2.
\]
Finite KL and $a>0$ imply $e_t>0$, using the usual extended-value convention at zero.  Consequently
\[
  e_t\ge\exp\{-(K_N(d)+\log2)/a\}
  \qquad(t\in T_N).
\]
Integrating over $T_N$, or summing with weight $\Delta_N$, and applying Lemma~\ref{lem:gp-budget} proves \eqref{supp-eq:gp-finite-budget}. If $R\ge a_\alpha d$, \eqref{supp-eq:gp-finite-lower} is immediate. Otherwise $a>a_\alpha$, and \eqref{supp-eq:gp-finite-budget} gives its second term.
\end{proof}

\section{Proof of the critical information rate}\label{supp-sec:gp-rate-proof}

Under the conditions of Corollary~\ref{cor:gp-rate}, an explicit constant in~\eqref{eq:gp-rate} is
\begin{equation}
\label{supp-eq:gp-rate-constant}
 c=a_\alpha
       \left(\frac{a_\alpha}{4\kappa C_0}\right)^{1/\kappa},
 \qquad a_\alpha=\frac{1-\alpha}{2}.
\end{equation}

\begin{proof}[Proof of Corollary~\ref{cor:gp-rate}]
Set $\ell_0=\operatorname{Leb}(J)>0$ and
\[
  r_N=(\log N/N)^{1/\kappa},\qquad
  \zeta=\left(\frac{a_\alpha}{4\kappa C_0}\right)^{1/\kappa},
  \qquad d_N=\zeta r_N.
\]
In the interval case, take $h_N=d_N$ and $T_N=J+h_N$. In the grid case, take
\[
  h_N=\Delta_N\lceil d_N/\Delta_N\rceil,
  \qquad T_N=(J\cap\Theta_N)+h_N.
\]
The mesh condition gives $h_N/d_N\to1$ on the grid, and the ratio equals one in the interval case.  In both cases $T_N\subseteq\Theta_N$, $h_N\le h_0$ eventually, the target mass $\ell_N$ is at least $\ell_0/2$ eventually, and each target $t=\theta+h_N$ has the partner $s=\theta$. For all sufficiently large $N$,
\[
  K_N(h_N)\le C_0Nh_N^\kappa
     \le \frac{a_\alpha}{2\kappa}\log N.
\]
Thus the exponential term in~\eqref{supp-eq:gp-finite-lower} is at least
\[
  \ell_N e^{-(K_N(h_N)+\log2)/a_\alpha}
  \ge \frac{\ell_0}{2}\,2^{-1/a_\alpha}N^{-1/(2\kappa)}.
\]
This lower bound divided by $r_N$ tends to infinity.  The grid correction $c_\alpha\Delta_N$ is $o(r_N)$, and $b_\alpha$ is fixed. Therefore the second term in~\eqref{supp-eq:gp-finite-lower} eventually exceeds $a_\alpha h_N$, uniformly over all honest procedures. It follows that $R_N(C)\ge a_\alpha h_N\ge a_\alpha\zeta r_N$. Taking the infimum proves~\eqref{eq:gp-rate} with the constant in~\eqref{supp-eq:gp-rate-constant}.
\end{proof}

\section{Roster conventions}\label{supp-sec:roster-conventions}

We use the notation and model of main Section~\ref{sec:roster-model}. Coverage is required for each fixed membership vector and every admissible response law. Procedures return subsets of the declared count range. Empty sets have diameter zero and count as coverage failures. Taking an integer hull preserves diameter. Probability and expectation include auxiliary randomisation when used.

\paragraph{Remark (Nonempty sets).} Fix a count $k^\dagger$ in the declared count range before observing responses, and replace every empty output by $\{k^\dagger\}$. The resulting set $\widetilde C$ satisfies
\[
\diam\widetilde C=\diam C\quad\text{on every outcome},\qquad
\Pp(K\in\widetilde C)\ge\Pp(K\in C).
\]
The expected-diameter rates can therefore also be attained with nonempty confidence sets.

\section{Statewise roster coverage and the baseline rate}\label{supp-sec:roster-baseline}

Use the fixed margins and notation of main Section~\ref{sec:roster-model}. Under statewise calibration, $|a_M(U)-r(U)|\le\varepsilon$ almost surely, and the reference and background restrictions in main~\eqref{eq:envclass} hold in every state. Let $\mathcal R^*_{n,m,\varepsilon,[b_L,b_U]}$ be the minimax expected count diameter over this class, allowing independent randomisation and full access to the indexed observations.
\begin{proposition}[Finite-sample coverage]\label{supp-thm:coverage}
For $k=0,\ldots,n$, define
\[
W_k=Y-kX/m,\quad
J_k=[(n-k)b_L-k\varepsilon,(n-k)b_U+k\varepsilon],
\]
\[
R_k=\sqrt{\frac{n+k^2/m}{2}\log\frac2\alpha},\qquad
C_H=\{k:\dist(W_k,J_k)\le R_k\}.
\]
Under conditional independence, $b_M(U)\in[b_L,b_U]$ and~\eqref{eq:class}, $\Pp(K\in C_H\mid U)\ge1-\alpha$, and hence $\Pp(K\in C_H)\ge1-\alpha$. The integer hull also covers. Empty sets are allowed, have diameter zero, and count as coverage failures. Their probability under any uniformly honest rule is therefore at most $\alpha$.
\end{proposition}
\begin{theorem}[Fixed-gap minimax expected diameter]\label{supp-thm:rate}
For all sufficiently large $n$, every integer $m\ge1$, and every admissible $\varepsilon,b_L,b_U$, constants $0<c<C<\infty$ depending only on the fixed margins and $\alpha$ satisfy
\[
c\min\{n,\sqrt n+n/\sqrt m+n\varepsilon+n\delta_b\}
\le\mathcal R^*_{n,m,\varepsilon,[b_L,b_U]}
\le C\min\{n,\sqrt n+n/\sqrt m+n\varepsilon+n\delta_b\}.
\]
The set $C_H\cap\mathcal K_n$ attains the upper rate. The deterministic weighted Chernoff/H refinement in Section~\ref{supp-sec:chernoff} also attains it. For $n\to\infty$, uniform vanishing relative expected diameter is possible exactly when $m\to\infty$, $\varepsilon\to0$, and $\delta_b\to0$ within this regime.
\end{theorem}

\subsection{Coverage and environmental mixtures}

\begin{proof}[Proof of Proposition~\ref{supp-thm:coverage}]
At $k=K$, $\E(W_K\mid U)=K(a_M-r)+(n-K)b_M\in J_K$. The independent conditional summands have squared range lengths summing to $n+K^2/m$. Hoeffding's bound \cite{hoeffding1963probability} gives each tail outside its true conditional mean a probability at most $\alpha/2$. Leaving the enlarged interval $J_K+[-R_K,R_K]$ requires one of those tail events. Apply a union bound to the two tails and integrate over $U$. The tail bound controls the exclusion of the true candidate, so inversion over the candidate set requires no multiplicity correction.
\end{proof}

\begin{proposition}[Mixture closure at fixed membership]\label{supp-prop:mix}
For each fixed vector $M$, let $\mathcal P_0(M)$ contain the independent Bernoulli product laws satisfying the statewise restrictions, and let $\mathcal P_{\rm mix}(M)$ allow arbitrary probability mixtures of these laws. For any fixed, possibly randomised procedure $C$, with independent auxiliary randomisation,
\begin{align*}
\sup_{P\in\mathcal P_{\rm mix}(M)}P\{K\notin C\}
&=\sup_{P\in\mathcal P_0(M)}P\{K\notin C\},\\
\sup_{P\in\mathcal P_{\rm mix}(M)}\E_P\diam C
&=\sup_{P\in\mathcal P_0(M)}\E_P\diam C.
\end{align*}
Consequently, taking the supremum over membership vectors gives identical uniformly honest procedure classes and identical minimax expected-diameter risks in the two experiments.
\end{proposition}
\begin{proof}
For $P=\int P_u\,dG(u)$, noncoverage probability and expected diameter are the corresponding integrals under $P_u$. Each integral is bounded by its component supremum. Conversely, point masses for $G$ include every independent component. These inequalities give equality for every procedure, first at fixed $M$, then uniformly over $M$. The resulting feasible procedure sets coincide, giving the same minimax risk. Randomisation can be integrated in the same way.
\end{proof}

Membership stays fixed throughout an environmental mixture. Under identical statewise restrictions, environmental mixing leaves the minimax risk unchanged.

\section{Roster rate and randomisation proofs}\label{supp-sec:roster-proofs}

We first establish the statewise and reliability rates, then prove the deterministic obstruction and the reference-randomness construction.

We use the fixed margins, declarations and count range of Theorem~\ref{supp-thm:rate}.
\begin{proof}[Upper bound in Theorem~\ref{supp-thm:rate}]
Let $b_c=(b_L+b_U)/2$ and $\widehat r=X/m$. Membership in $C_H$ is equivalent to
\[
|Y-nb_c-k(\widehat r-b_c)|
\le k\varepsilon+(n-k)\delta_b/2+R_k.
\]
Subtracting the residuals for two retained candidates gives
\begin{equation}\label{supp-eq:diam}
\diam(C_H)|\widehat r-b_c|\le2n\varepsilon+n\delta_b+2R_n.
\end{equation}
Let $g_0=r_--\bar b>0$. On $|\widehat r-r(U)|\le g_0/2$, the denominator is at least $g_0/2$. The complementary event has probability at most $2e^{-mg_0^2/2}$, by the conditional reference Hoeffding bound. Hence uniformly,
\[
\E\diam C_H\le\min\left\{n,
\frac{4n\varepsilon+2n\delta_b+4R_n}{g_0}+2ne^{-mg_0^2/2}\right\}.
\]
Now $R_n\le\sqrt{\log(2/\alpha)/2}(\sqrt n+n/\sqrt m)$ and $e^{-cm}\le C_c/\sqrt m$ for $m\ge1$. Coverage follows from Proposition~\ref{supp-thm:coverage}. Intersecting with the fixed count range preserves coverage and can only reduce diameter.
\end{proof}

\subsection{Lower bounds from the complete observation law}\label{supp-sec:roster-baseline-lower}

For two laws $P_0,P_1$ with different counts $k_0,k_1$, any honest procedure satisfies
\begin{equation}\label{supp-eq:two}
\E_{P_0}\diam C\ge |k_1-k_0|
\{1-2\alpha-\TV(P_0,P_1)\}_+.
\end{equation}
Indeed, $P_0(k_0\in C)\ge1-\alpha$ and $P_0(k_1\in C)\ge1-\alpha-\TV(P_0,P_1)$. Both counts must be retained on their intersection. All constructions below use a degenerate environment, as permitted by Proposition~\ref{supp-prop:mix}. The bounds apply to procedures using the full unit-indexed response vector.

For main sampling noise, embed $b=b_L$ and exact calibration with a fixed $r_0\in(r_-,r_+)$. At each count $k$, assign member probability $r_0$ and nonmember probability $b$, and average the laws uniformly over membership permutations. Denote this proof mixture by $Q_k$. Conditional on $Y$, the unit-indexed main response vector is uniform over binary vectors with that total. Its conditional law is the same for every $k$, and the reference law is also unchanged. Thus the TV of full observations equals the TV of the count distributions.

For interior $k$ and $k+h$, the totals have common part $V\sim\operatorname{Bin}(k,r_0)+\operatorname{Bin}(n-k-h,b)$ and an additional block of $h$ Bernoulli trials with probability $b$ or $r_0$. Use common uniforms for the two additional blocks, independently of $V$, and denote their totals by $A_b,A_r$. Then $A_r\ge A_b$ and $\E(A_r-A_b)=h(r_0-b)$. Mixture convexity, translation invariance and telescoping give
\[
\begin{aligned}
\TV(Q_k,Q_{k+h})
&=\TV(V+A_b,V+A_r)\\
&\le\E\{\TV(V,V+A_r-A_b)\}\\
&\le\E(A_r-A_b)\,\TV(V,V+1)\\
&=h(r_0-b)\TV(V,V+1)
\le h(r_0-b)\TV(B,B+1),
\end{aligned}
\]
where $B\sim\operatorname{Bin}(k,r_0)$ and the last step is convolution contraction. Binomial unimodality gives $\TV(B,B+1)=\max_j P(B=j)\le C/\sqrt{k r_0(1-r_0)}$. One direct bound follows by integrating the characteristic-function envelope $\exp\{-2k r_0(1-r_0)\sin^2(t/2)\}$ over $[-\pi,\pi]$. Consequently TV is at most $C'h/\sqrt n$. Choose $h=\lfloor c_1\sqrt n\rfloor$ with small $c_1$ and use~\eqref{supp-eq:two}. Uniform fixed-vector coverage transfers to each proof mixture, whose expected diameter is bounded by the largest fixed-vector risk.

For reference sampling noise, choose $k_0/n$ strictly between $\rho_-$ and $\rho_+$ and fix $r_0\in(r_-,r_+)$. Hold every main probability fixed at
\[
q_i=r_0\quad(i\le k_0),\qquad q_i=b_L\quad(i>k_0).
\]
Model 0 assigns membership to the first $k_0$ positions and uses reference probability $r_0$. Model 1 assigns the first $k_1=k_0+h$ positions to membership and uses
\[
r_1=b_L+\frac{k_0(r_0-b_L)}{k_0+h}
\]
for every reference. Both models satisfy exact group-average calibration. The entire main law is identical, even though Model 1 has heterogeneous members. Only the reference law distinguishes them. With $h/n$ sufficiently small, both reference means stay inside the fixed margins, and
\[
\KL(P_0,P_1)=m\,\mathrm{kl}(r_0,r_1)
\le\frac{m(r_0-r_1)^2}{r_1(1-r_1)}\le C m h^2/n^2.
\]
Pinsker's inequality and $h=\lfloor c_2n/\sqrt m\rfloor$ give TV bounded away from $1-2\alpha$ for sufficiently small fixed $c_2$. For $m\le n$, rounding is harmless at large $n$, and~\eqref{supp-eq:two} gives the claimed term. For $m>n$, it is already dominated by $\sqrt n$. Constants can be chosen uniformly in $b_L\in[0,\bar b]$.

For calibration ambiguity, keep the same main vector $q_i=r_0$ for $i\le k_0$ and $q_i=b_L$ otherwise, and now keep \emph{both} reference laws at $r_0$. Again change membership from the first $k_0$ to the first $k_0+h$ positions. The second member average is
\[
a_1=b_L+\frac{k_0(r_0-b_L)}{k_0+h},\qquad
|a_1-r_0|=\frac{h(r_0-b_L)}{k_0+h}.
\]
Take $h=\lfloor c_3n\varepsilon\rfloor$ for a small fixed $c_3$. The counts stay interior and this discrepancy is at most $\varepsilon$. The two membership assignments induce identical full observation laws. Equation~\eqref{supp-eq:two} has TV zero. When $n\varepsilon$ exceeds a sufficiently large multiple of $\sqrt n$, rounding leaves a constant fraction of $n\varepsilon$. At smaller values, the main-noise lower bound absorbs this term.

For background ambiguity, choose $k_H/n$ strictly inside the count range and set $q_i=r_0$ for $i\le k_H$, $q_i=b_L$ otherwise. All references have probability $r_0$ in both models. In the high-count model, the first $k_H$ positions are members, and the nonmember mean is $b_L$. In the low-count model, only the first $k_L=k_H-h$ positions are members. Its member mean is still exactly $r_0$, while its nonmember mean is
\begin{equation}\label{supp-eq:bpair}
b'=b_L+\frac{h(r_0-b_L)}{n-k_H+h}.
\end{equation}
For $h=\lfloor c_4 n\delta_b\rfloor$, choose $c_4$ small enough that both counts remain interior and $c_4r_0/(1-k_H/n)\le1$. Such a choice is uniform at large $n$ because $k_H/n$ is bounded away from 1. Then $b'\le b_L+\delta_b=b_U$. Calibration is exact and the full main/reference law is identical. The TV-zero bound yields $\Omega(n\delta_b)$ whenever that term dominates rounding. At smaller values, $\sqrt n$ absorbs it.

Every honest procedure faces all four submodels. The maximum of four nonnegative terms is at least their sum divided by four. The interior count range has length proportional to $n$, and all four terms are at most constant multiples of $n$. Combining the bounds gives the asserted clipped rate. Dividing by $n$ proves the stated necessary and sufficient conditions for vanishing relative expected diameter.

We use the reliability class in~\eqref{eq:envclass} and the quantities $s,u,v$ from Theorem~\ref{thm:environment}. Write $L(s,u)=\log[\mathrm e/\min\{1,s+u\}]$.

\subsection{Finite-sample construction and upper bound}\label{supp-env:sec:upper}

Let $X=\sum_j V_j$ and $Y=\sum_iZ_i$. For $0<\beta<1$, define
\[
W_k=Y-kX/m,\quad
J_k=[(n-k)b_L-k\varepsilon,(n-k)b_U+k\varepsilon],\quad
R_k(\beta)=\sqrt{\frac{n+k^2/m}{2}\log\frac2\beta},
\]
and
\[
H_\beta=\{k\in\mathcal K_n:\dist(W_k,J_k)\le R_k(\beta)\}.
\]
On every good state, the conditional mean of $W_K$ belongs to $J_K$. The independent summands have squared range lengths summing to $n+K^2/m$. Hoeffding's two one-sided bounds give conditional noncoverage probability at most $\beta$. Since the actual bad-state mass is at most the declared $\eta$,
\begin{equation}\label{supp-env:eq:raw}
\Pp(K\notin H_\beta)\le q_\beta:=\eta+(1-\eta)\beta.
\end{equation}
For $\eta<\alpha$, the familiar allocation $\beta=(\alpha-\eta)/(1-\eta)$ is already honest without any extra randomisation. This allocation degenerates at the critical point.

For all $\eta$, let
\begin{equation}\label{supp-env:eq:coin}
t_\beta=\left(1-\frac{\alpha}{q_\beta}\right)_+.
\end{equation}
Independently of the responses, return $\mathcal K_n$ with probability $t_\beta$, and otherwise return the integer hull of $H_\beta$. Denote this set by $C_\beta$. Its noncoverage probability is at most $(1-t_\beta)q_\beta\le\alpha$ at every finite sample size. The full-range probability is specified before observing the responses.

Put $b_c=(b_L+b_U)/2$ and $\widehat r=X/m$. For any observed data, two retained candidates in $H_\beta$ satisfy
\begin{equation}\label{supp-env:eq:residual}
\diam H_\beta\,|\widehat r-b_c|
\le2n\varepsilon+n\delta_b+2R_n(\beta).
\end{equation}
This is an algebraic property of the inverted set, even when the true count does not satisfy the calibration declaration. Let $g_0=r_--\bar b>0$. Because the reference restriction holds in every state,
\[
\Pp\{|\widehat r-r(U)|>g_0/2\}\le2\exp(-mg_0^2/2).
\]
On the complementary event $|\widehat r-b_c|\ge g_0/2$. Thus, with $D_n=\diam\mathcal K_n$,
\begin{equation}\label{supp-env:eq:Hupper}
\E\diam H_\beta\le B_\beta:=
\min\left\{D_n,\frac{4n\varepsilon+2n\delta_b+4R_n(\beta)}{g_0}
+2n e^{-mg_0^2/2}\right\}.
\end{equation}
Consequently the explicit finite-sample bound is
\begin{equation}\label{supp-env:eq:Cupper}
\E\diam C_\beta\le t_\beta D_n+(1-t_\beta)B_\beta.
\end{equation}
The all-state reference separation controls diameter in both good and bad states.

Use $s,u$ from Theorem~\ref{thm:environment} and choose $\beta=(u+s)/4$. This lies in $(0,3/8)$. If $\eta\le\alpha$,
\[
q_\beta-\alpha=-u+(1-\eta)(u+s)/4\le s/4;
\]
if $\eta>\alpha$, then $q_\beta-\alpha\le v+s/4$. Whenever $t_\beta>0$, $q_\beta\ge\alpha$, so in both cases
\[
t_\beta\le(v+s/4)/\alpha.
\]
For $s<1$, $R_n(\beta)/n\le C s\sqrt{\log(2/\beta)}$ and $e^{-mg_0^2/2}\le C m^{-1/2}\le Cs$. Also $\log(2/\beta)=\log(8/(u+s))\le C L(s,u)$, since $L\ge1$. Apply~\eqref{supp-env:eq:Cupper}, absorb $s$ into $s\sqrt L$, and clip at $n$. For $s=1$ the trivial roster set gives the same upper order. This proves the upper half of~\eqref{eq:envrate} uniformly in $\eta$.

\subsection{Common bad states and a sharper many-count bound}\label{supp-env:sec:common}

Fix $b=b_L$ and $r_0=(r_-+r_+)/2$. For every $k\in\mathcal K_n$, the same full response law
\[
Q:\quad q_i=b\ (\text{all main units}),\qquad p_j=r_0\ (\text{all references})
\]
is an admissible bad state when the first $k$ units are members. Its mismatch is $r_0-b>\bar\varepsilon$. Choose any exactly calibrated admissible good law $G_k$ for that fixed membership vector, and form
\begin{equation}\label{supp-env:eq:mixture}
P_k=(1-\eta)G_k+\eta Q.
\end{equation}
This is a two-state shared-environment model. Both component laws are conditionally independent product laws with the same fixed membership vector. The observed response law $Q$ is common to all the membership interpretations.

\begin{proposition}[The many-count budget inequality]\label{supp-env:prop:mass}
For any procedure $C$ honest over $\mathcal P_\eta$ with worst-case expected diameter $R$, and any integer interval $I\subseteq\mathcal K_n$ for which the mixtures~\eqref{supp-env:eq:mixture} are available, put $N_I=|I|$. Then
\begin{equation}\label{supp-env:eq:mass}
R\ge\big\{(\eta-\alpha)N_I-\eta\big\}_+.
\end{equation}
If $\eta<1$, the good-state errors $e_k=G_k(k\notin C)$ also obey
\begin{equation}\label{supp-env:eq:averageerror}
\frac1{N_I}\sum_{k\in I}e_k
\le\frac{u+(R+\eta)/N_I}{1-\eta},\qquad
\E_{G_k}\diam C\le\frac{R}{1-\eta}.
\end{equation}
\end{proposition}
\begin{proof}
Write $p_k=Q(k\in C)$. Honesty gives
\[
(1-\eta)e_k+\eta(1-p_k)\le\alpha.
\]
Pointwise $|C\cap I|\le\diam C+1$, including an empty set. Therefore $\sum_{k\in I}p_k\le\E_Q\diam C+1$. Nonnegative risks in the mixture imply $\eta\E_Q\diam C\le R$. Summing the coverage inequalities first after dropping the nonnegative good-state error gives~\eqref{supp-env:eq:mass}. Retaining it and using $\alpha-\eta\le u$ gives~\eqref{supp-env:eq:averageerror}. The last risk inequality follows directly from~\eqref{supp-env:eq:mixture}. The argument includes independent randomisation by enlarging the common law with the same independent seed.
\end{proof}

Summing over a number of counts proportional to $n$ places the threshold at $\eta=\alpha$, up to the integer correction in~\eqref{supp-env:eq:mass}. Equation~\eqref{supp-env:eq:averageerror} requires small good-state errors at most counts when the set is narrow and $\eta$ is near $\alpha$.

\subsection{Smooth families for the full unit-indexed observation law}\label{supp-env:sec:families}

We construct two families of laws of the full unit-indexed response vector with fixed membership vectors. Put $\rho_0=(\rho_-+\rho_+)/2$ and choose
\[
\ell=\frac1{16}\min\left\{\rho_+-\rho_0,\rho_0,
\frac{\rho_0(1-r_0)}{r_0},\frac{\rho_0(r_0-r_-)}{r_0}\right\}>0.
\]
Let $k_*=\lceil\rho_0 n\rceil$ and $I=\{k_*,\ldots,k_*+\lfloor\ell n\rfloor\}$. For sufficiently large $n$, this lies inside $\mathcal K_n$ and $N_I\ge\ell n$. In every model indexed by $k\in I$, the first $k$ main units are the fixed members and $b=b_L$.

\paragraph{Main-response family.} Set
\[
q_i(k)=\begin{cases}b+\dfrac{k}{k_*}(r_0-b),&i\le k_*,\\b,&i>k_*,\end{cases}
\qquad p_j(k)=r_0.
\]
Its member average is exactly $r_0$, its nonmember average is $b$, and its high probability lies in a fixed interior interval between $r_0$ and $r_0+(1-r_0)/16<1$. Only the first $k_*$ response probabilities change with $k$. For $j,k\in I$, the product KL is bounded by
\begin{equation}\label{supp-env:eq:KLmain}
\KL(G_k^{\mathrm{main}},G_j^{\mathrm{main}})\le C\frac{(j-k)^2}{n}.
\end{equation}
Indeed, $\mathrm{kl}(p,q)\le(p-q)^2/[q(1-q)]$, the denominator is bounded below, and the probability change is $(j-k)(r_0-b)/k_*$.

\paragraph{Reference-response family.} Set
\[
q_i(k)=\begin{cases}r_0,&i\le k_*,\\b,&i>k_*,\end{cases}
\qquad p_j(k)=r(k)=b+\frac{k_*}{k}(r_0-b).
\]
The main law is fixed. The member average equals $r(k)$ and the nonmember average equals $b$. The choice of $\ell$ ensures $r(k)\ge r_0-(r_0-r_-)/16>r_-$, and $r(k)\le r_0<r_+$. Since $|r(j)-r(k)|\le C|j-k|/n$,
\begin{equation}\label{supp-env:eq:KLref}
\KL(G_k^{\mathrm{ref}},G_j^{\mathrm{ref}})\le C m\frac{(j-k)^2}{n^2}.
\end{equation}
All constants are uniform over the declared background endpoints. At zero background probability, the background coordinates have identical laws throughout the family, and the changing probabilities retain fixed interior margins. Either family can be combined with the common bad state $Q$ in Section~\ref{supp-env:sec:common}.

\subsection{Lower bound, including the critical logarithm}\label{supp-env:sec:lower}

We first record the ordinary four-term lower bound, then use the preceding families and the average good-state error constraint to strengthen its sampling terms.

\subsubsection{A baseline bound uniform over calibration-failure bounds}

Since $\mathcal P_0\subseteq\mathcal P_\eta$, Theorem~\ref{supp-thm:rate} implies, uniformly over $\eta$,
\begin{equation}\label{supp-env:eq:baseline}
R\ge c n\min\{1,s+\varepsilon+\delta_b\}.
\end{equation}
The many-count budget inequality~\eqref{supp-env:eq:mass}, used on $\mathcal K_n$, additionally gives $R\ge c'nv-1$. The integer correction is absorbed by~\eqref{supp-env:eq:baseline}. It remains to establish the strengthened sampling term near the reliability boundary.

\subsubsection{Small diameter forces high good-state confidence}

It remains to prove the sampling lower bound $R\ge c n s\sqrt{L(s,u)}$. We can restrict to
\[
\alpha/2\le\eta\le(1+\alpha)/2.
\]
If $\eta<\alpha/2$, then $u\ge\alpha/2$ and $L$ is bounded, so~\eqref{supp-env:eq:baseline} suffices. If $\eta>(1+\alpha)/2$, then $v>(1-\alpha)/2$ and the common-state bound is of order $n$, which also suffices. Within the displayed range, $1-\eta$ is bounded below by a fixed positive constant.

Write $w=R/n$. By~\eqref{supp-env:eq:baseline}, $w\ge c s$ and $R\ge c\sqrt n$. Use either smooth family on $I$. From~\eqref{supp-env:eq:averageerror}, its mean good-state error is at most
\[
C_0(u+w+1/n).
\]
At least $7N_I/8$ of its counts therefore have error at most
\[
\beta_0=8C_0(u+w+1/n).
\]
The constants here and below depend only on the fixed margins and $\alpha$. We may restrict to $u+w$ smaller than a sufficiently small fixed constant. Otherwise either $u$ is bounded below, making $L$ bounded and~\eqref{supp-env:eq:baseline} sufficient, or $w$ is bounded below and the desired lower bound is automatic because $s\sqrt L\le1$.

Choose the integer shift
\[
h=\left\lceil\frac{8R}{1-\eta}\right\rceil+1.
\]
By reducing that fixed smallness threshold, $h\le N_I/4$ and $\beta_0$ is as small as needed. Among the $N_I-h$ pairs $k,k+h$ in $I$, at most $N_I/4$ have a bad endpoint, so there exists a pair with both good-state errors at most $\beta_0$. Under $G_k$,
\[
G_k\{k\in C,\diam C<h\}
\ge1-\beta_0-\frac{\E_{G_k}\diam C}{h}
\ge1-\beta_0-\tfrac18\ge\tfrac34.
\]
On this event $k+h\notin C$. Under $G_{k+h}$, that exclusion has probability at most $\beta_0$. Data processing to this binary event gives
\begin{equation}\label{supp-env:eq:binaryKL}
\KL(G_k,G_{k+h})\ge\tfrac34\log(1/\beta_0)-\log2.
\end{equation}
The same statement holds for randomised procedures because appending a common independent seed leaves KL unchanged.

Since $1/n\le Cw$ and $h\le CR$ for large $n$, combine~\eqref{supp-env:eq:binaryKL} with~\eqref{supp-env:eq:KLmain} or~\eqref{supp-env:eq:KLref}. If necessary reduce the smallness threshold once more to absorb fixed logarithmic constants. This yields, respectively,
\[
w\ge c n^{-1/2}\sqrt{\log\frac{\mathrm e}{u+w}},\qquad
w\ge c m^{-1/2}\sqrt{\log\frac{\mathrm e}{u+w}}.
\]
Taking their maximum and using $s\le2\max\{n^{-1/2},m^{-1/2}\}$ gives
\begin{equation}\label{supp-env:eq:implicit}
w\ge c s\sqrt{\log\frac{\mathrm e}{u+w}}.
\end{equation}

\subsubsection{Removing the implicit risk from the logarithm}

If $s+u\ge1$, $L(s,u)=1$ and the baseline bound finishes the proof. Otherwise write $L=\log(\mathrm e/(s+u))\ge1$ and $f=s\sqrt L\le1$. If $w\ge f$, there is nothing to prove. If $w<f$, then
\[
u+w\le u+s\sqrt L\le(u+s)\sqrt L,
\quad\text{so}\quad
\log\frac{\mathrm e}{u+w}\ge L-\tfrac12\log L\ge\tfrac12 L.
\]
Equation~\eqref{supp-env:eq:implicit} gives $w\ge c s\sqrt L$. The cases previously disposed of satisfy the same conclusion by constants depending only on the fixed margins and $\alpha$. Combining this sampling bound,~\eqref{supp-env:eq:baseline}, and~\eqref{supp-env:eq:mass}, using that a maximum of finitely many nonnegative clipped terms dominates their clipped sum up to a constant, proves the lower half of Theorem~\ref{thm:environment}.

The phase statements follow directly. For a fixed subcritical declaration $\eta$, $u>0$ makes $L$ bounded. At equality, $u=0$ gives the stated logarithm. For fixed supercritical $\eta$, $v>0$ gives order $n$. Along general sequences, $s\sqrt{\log(\mathrm e/s)}\to0$ if $s\to0$, and $s\sqrt{L(s,u)}$ lies between $s$ and that expression. Thus~\eqref{eq:envrate} tends to zero after division by $n$ exactly when $m\to\infty$, $\varepsilon,\delta_b\to0$ and $(\eta-\alpha)_+\to0$.\hfill$\square$

\subsection{Deterministic procedures}

\begin{proposition}[Finite certificate for the deterministic obstruction]\label{supp-det:finite}
In the unrestricted-probability reliability class, suppose $\eta\ge\alpha$ and the background is the known rational number $b_L=b_U=b\in[0,\bar b]$. Let $d_b$ be its denominator in lowest terms, with $d_b=1$ when $b=0$. Write $k_-=\min\mathcal K_n$, $k_+=\max\mathcal K_n$, $D_n=k_+-k_-$, and
\[
r_0=(r_-+r_+)/2,\qquad d_r=\min\{r_0-r_-,r_+-r_0\},\qquad c_r=\min\{r_0,1-r_0\}.
\]
Let $G_-$ be the pure clean product law with $K=k_-$, member and reference probabilities $r_0$, and nonmember probabilities $b$. Every deterministic uniformly honest confidence set satisfies
\begin{equation}\label{supp-eq:detpoint}
\E_{G_-}\diam C\ \ge\ (D_n-2d_b)_+
\left[1-2e^{-2md_r^2}-2e^{-2k_-^2c_r^2/n}\right]_+.
\end{equation}
The same lower bound applies to $\mathcal R^*_{\eta,\mathrm{det}}$, which is at most $D_n$. For $b=0$, the stronger bound is
\begin{equation}\label{supp-eq:detzero}
\E_{G_-}\diam C\ge D_n\Pp\{\mathrm{Bin}(m,r_0)/m\in[r_-,r_+]\}.
\end{equation}
Consequently, along $n,m\to\infty$ with $d_{b_n}=o(n)$ for a sequence $b=b_n$,
\[
\mathcal R^*_{\eta,\mathrm{det}}/D_n\longrightarrow1.
\]
This conclusion holds even when $\varepsilon=0$. At $\eta=\alpha$, $m\asymp n$, $\varepsilon=0$ and a fixed rational background, the randomised minimax order is $\sqrt{n\log n}$, whereas the deterministic minimax diameter is asymptotic to $D_n$.
\end{proposition}

We prove Theorem~\ref{thm:deterministic} using its notation.
\begin{lemma}[An atom that exhausts the confidence budget]\label{supp-lem:atom}
Let $C$ be deterministic and uniformly honest over the unrestricted-probability reliability class with $\eta\ge\alpha$. Fix a response vector $w=(z,v)$ and a candidate count $k$. Suppose that, for some fixed membership vector $M$ of size $k$, the Dirac product state $\delta_w$ satisfies the all-state reference and background restrictions. If it is a good state, then $k\in C(w)$. If it is a bad state and there is an admissible good product law $G_M$ with $G_M(w)>0$, the same conclusion holds.
\end{lemma}
\begin{proof}
Excluding $k$ at a good Dirac state gives noncoverage probability one under an admissible pure law. For a bad Dirac state, the two-state law
\[
P_M=(1-\alpha)G_M+\alpha\delta_w
\]
belongs to the class because its bad-state mass is $\alpha\le\eta$ and its membership vector is the same in both states. If $k\notin C(w)$, determinism implies
\[
P_M(k\notin C)\ge\alpha+(1-\alpha)G_M(w)>\alpha,
\]
contradicting honesty. The strict inequality follows from $G_M(w)>0$.
\end{proof}

For $x=\sum_jv_j$ and $y=\sum_iz_i$, assume $x/m\in[r_-,r_+]$. A Dirac product state at $w$ can have count $k$ and background $b$ if and only if
\begin{equation}\label{supp-eq:atomcompat}
(n-k)b\in\mathbb Z,\qquad (n-k)b\le y\le k+(n-k)b.
\end{equation}
Necessity follows because the $n-k$ nonmembers must contain exactly $(n-k)b$ observed ones. For sufficiency, place that many ones and $(n-k)(1-b)$ observed zeros among the nonmembers. The two inequalities guarantee enough ones and zeros, and all remaining positions are members. Set the individual probabilities to the observed zeros and ones. The resulting full vector is a conditionally independent Bernoulli product state, with reference mean $x/m$ and background mean $b$.

For this fixed assignment $M$, let $G_M$ have homogeneous member and reference probabilities $r_0$ and nonmember probabilities $b$. If $b>0$, every coordinate probability is strictly between zero and one, so $G_M(w)>0$. If $b=0$, compatibility assigns only observed zeros to nonmembers. These have probability one, and all other coordinates have interior probabilities. Thus $G_M(w)>0$ also in that case. This law is exactly calibrated, hence good for every admissible $\varepsilon$. The two branches of Lemma~\ref{supp-lem:atom} cover the good and bad Dirac states. Therefore
\begin{equation}\label{supp-eq:forcedset}
\{k\in\mathcal K_n:\text{\eqref{supp-eq:atomcompat} holds}\}\subseteq C(w)
\quad\text{whenever }x/m\in[r_-,r_+].
\end{equation}
Different candidates may require different membership assignments. Uniform honesty is imposed on each such fixed-assignment experiment separately. Membership remains fixed across environmental states, and procedures may use the full unit-indexed response vector.

The integrality condition is $d_b\mid(n-k)$. Define
\[
I_b=\{k\in\mathcal K_n:d_b\mid(n-k)\}.
\]
The diameter of this arithmetic progression is at least $(D_n-2d_b)_+$ (zero if empty). For every $k\ge k_-$, the feasible interval in~\eqref{supp-eq:atomcompat} contains
\[
[(n-k_-)b,\ k_-+(n-k_-)b],
\]
because its lower endpoint decreases and its upper endpoint increases with $k$. Thus all of $I_b$ is forced into $C$ on the event
\begin{equation}\label{supp-eq:goodwindow}
X/m\in[r_-,r_+],\qquad (n-k_-)b\le Y\le k_-+(n-k_-)b.
\end{equation}
Under $G_-$, $X\sim\mathrm{Bin}(m,r_0)$ and $\E Y=k_-r_0+(n-k_-)b$. The distances from this mean to the two main-total endpoints are $k_-r_0$ and $k_-(1-r_0)$. Hoeffding's inequality and a union bound imply that~\eqref{supp-eq:goodwindow} has probability at least
\[
\left[1-2e^{-2md_r^2}-2e^{-2k_-^2c_r^2/n}\right]_+.
\]
Multiplying by the forced lattice diameter proves~\eqref{supp-eq:detpoint}. Since $G_-$ is itself in the class, this is also a minimax lower bound. The procedure returning $\mathcal K_n$ always is honest, giving the upper bound $D_n$.

If $b=0$, every integer candidate lies in $I_b$, and under $G_-$ the main total obeys $0\le Y\le k_-$ almost surely. On the reference-window event the entire $\mathcal K_n$ is therefore forced, giving~\eqref{supp-eq:detzero}. Finally $D_n/n\to\rho_+-\rho_->0$, $k_-/n\to\rho_->0$, and both concentration errors vanish when $n,m\to\infty$. If $d_{b_n}=o(n)$, the lattice loss also vanishes relative to $D_n$. Squeezing between this bound and $D_n$ proves the asserted limit. The randomised comparison follows from the matching environmental theorem.\hfill$\square$

\begin{corollary}[Other favourable counts]\label{supp-cor:othercounts}
Retain the assumptions and notation of Theorem~\ref{thm:deterministic}. For any $k_0\in\mathcal K_n$, let $G_{k_0}$ have member and reference probabilities $r_0$, nonmember probability $b$, and true count $k_0$. Put
\[
\lambda_b=\frac{r_0-b}{1-b},\qquad z_{k_0}=k_0(r_0-b)+k_-b.
\]
For every $t>0$, each deterministic uniformly honest rule satisfies
\[
\begin{split}
\E_{G_{k_0}}\diam C\ \ge{}&
\left[k_+-\max\left\{k_-,\lambda_b k_0+\frac{t}{1-b}\right\}-2d_b\right]_+\\
&\times\left[1-2e^{-2md_r^2}-e^{-2t^2/n}-e^{-2z_{k_0}^2/n}\right]_+.
\end{split}
\]
For fixed rational $b$, $n,m\to\infty$ and $k_0/n\to\kappa_0\in[\rho_-,\rho_+]$, this implies
\[
\liminf\frac{\E_{G_{k_0}}\diam C}{D_n}
\ge\frac{\rho_+-\max\{\rho_-,\lambda_b\kappa_0\}}{\rho_+-\rho_-}.
\]
More generally, for any rational-background sequence $b=b_n\in[0,\bar b]$ with $d_{b_n}=o(n)$, the following bound holds uniformly in $k_0$.
\[
\inf_{k_0\in\mathcal K_n}\frac{\E_{G_{k_0}}\diam C}{D_n}
\ge c_*+o(1),\qquad
c_*:=\frac{\rho_+-\max\{\rho_-,r_0\rho_+\}}{\rho_+-\rho_-}>0.
\]
\end{corollary}
\begin{proof}
Under $G_{k_0}$, $\mu:=\E Y=nb+k_0(r_0-b)$ and $X\sim\operatorname{Bin}(m,r_0)$. On the event
\[
X/m\in[r_-,r_+],\qquad (n-k_-)b\le Y\le\mu+t,
\]
every candidate with $d_b\mid(n-k)$ and $k\ge\max\{k_-,\lambda_b k_0+t/(1-b)\}$ satisfies~\eqref{supp-eq:atomcompat}. Its lower compatibility inequality follows from $k\ge k_-$, and its upper inequality follows from $Y\le nb+k_0(r_0-b)+t\le nb+k(1-b)$. Equation~\eqref{supp-eq:forcedset} retains this lattice. Rounding its two endpoints loses at most $2d_b$ in diameter. The lower-tail distance for $Y$ is $\mu-(n-k_-)b=z_{k_0}$, so Hoeffding's inequality and a union bound give the displayed probability factor.

Take $t=\sqrt{n\log n}$. Then $t/n\to0$, $z_{k_0}\ge k_-r_0$, and all probability errors vanish as $n,m\to\infty$. Division by $D_n$ proves the fixed-$b$ limit. For varying backgrounds, $1-b\ge1-\bar b>0$ and $\lambda_b\le r_0<1$ uniformly. The same calculation, with $d_{b_n}/n\to0$, gives $c_*$.
\end{proof}
For $(\rho_-,\rho_+,r_0,b)=(0.2,0.7,0.5,0.1)$, $\lambda_b=4/9$. The limiting lower-bound fractions are $1$, $1$ and $7/9$ at $k_0/n=0.2$, $0.4$ and $0.7$, respectively. The corollary therefore gives a positive fraction throughout the count range and asymptotically full-range expected diameter where $\lambda_b\kappa_0\le\rho_-$.

\subsection{Subcritical calibration}

For any fixed $\eta<\alpha$, set $\beta=(\alpha-\eta)/(1-\eta)>0$ in the deterministic inversion $H_\beta$. Its total error is at most $\eta+(1-\eta)\beta=\alpha$. The same all-state reference separation bounds its expected diameter by the original four-term order. The randomised lower bound also applies to the smaller deterministic procedure class, so the two minimax orders agree for every fixed subcritical $\eta$, with constants that may depend on its distance from $\alpha$.

Theorem~\ref{thm:deterministic} uses compatible Dirac response states and a rational background. Corollary~\ref{supp-cor:backgroundinterval} extends the background condition to intervals. Theorem~\ref{thm:interior} gives the corresponding rate when a prespecified subset of reference responses has uniformly interior conditional probabilities.
\begin{corollary}[Background intervals containing a short-denominator rational]\label{supp-cor:backgroundinterval}
Retain the unrestricted individual probabilities and $\eta\ge\alpha$ of Theorem~\ref{thm:deterministic}, but allow a declared interval $[b_{L,n},b_{U,n}]$. If it contains a rational $b_n$ with reduced denominator $d_{b_n}$, the pointwise bound~\eqref{supp-eq:detpoint} holds at the pure clean law with background $b_n$. Hence $\mathcal R^*_{\eta,\mathrm{det}}/D_n\to1$ whenever $n,m\to\infty$ and $d_{b_n}=o(n)$. A sufficient condition for positive-width intervals is
\[
n\delta_{b,n}\to\infty.
\]
At $\eta=\alpha$, $m\asymp n$, $\varepsilon=0$ and $\delta_{b,n}=n^{-1/2}$, the randomised rate is $\sqrt{n\log n}$ while the deterministic risk is asymptotic to $D_n$.
\end{corollary}
\begin{proof}
An honest procedure for an interval declaration must be honest for every singleton submodel it contains. Apply Theorem~\ref{thm:deterministic} at $b_n$. Its reference and main concentration factors do not depend on $b_n$, so the same conclusion holds for a varying sequence of rationals. For an interval of width $\delta>0$, choose $Q=\lceil1/\delta\rceil$ and $p=\lceil Qb_L\rceil$. Then $b_L\le p/Q<b_L+1/Q\le b_U$. Reducing the fraction gives a denominator $d_b\le Q$. Thus
\[
\frac{d_b}{n}\le\frac1{n\delta}+\frac1n\longrightarrow0.
\]
In particular a finite-sample lower certificate for any such interval is~\eqref{supp-eq:detpoint} with $d_b$ replaced by $Q$. An arbitrarily narrow interval containing a fixed rational also satisfies the corollary. The critical-rate example follows from Theorem~\ref{thm:environment} because $n\delta_{b,n}=\sqrt n$ is absorbed by $\sqrt{n\log n}$.
\end{proof}

\subsection{Randomness from the reference sample}\label{supp-sec:reserve_construction}

We specify the matrices and reserve sizes used in Theorem~\ref{thm:interior}. Retain its fixed margins and known $0<\tau<1/2$ with $\tau<r_-<r_+<1-\tau$. Put $g_0=r_--\bar b$. For $m\ge2$, set $d_m=\lceil\log_2m\rceil$, $\rho=1-2\tau$ and $c_\tau=(1+\rho^2)/2$. Fix a binary $d_m\times h_m$ matrix $A_m$ satisfying
\[
h_m\le\left\lceil\frac{\log\{2m^2(2^{d_m}-1)\}}{-\log c_\tau}\right\rceil,
\qquad
\sum_{u\in\mathbb F_2^{d_m}\setminus\{0\}}\rho^{2w(A_m^\top u)}\le m^{-2},
\]
where $w$ is Hamming weight. Lemma~\ref{supp-lem:exist} proves existence with $h_m=O_\tau(\log m)$. The matrix and its length are fixed before responses are observed. Use the reserve when $h_m\le m/2$ and $h_m/m\le g_0/2$. At other sizes, return $\mathcal K_n$, as specified below.
\begin{lemma}[Transfer from conditional seed approximation]\label{supp-lem:internal}
Let $T$ be retained inference data and $J$ a deterministic function of reserved data. Suppose, uniformly over the model and every state $U$, that $J$ is independent of $T$ conditional on $U$ and
\[
\TV\{\mathcal L(J\mid U),\nu\}\le\Delta_{\rm seed},
\]
where $\nu$ is a fixed probability law on a finite seed space. For a rule $F(T,j)\subseteq\mathcal K_n$, let $J^\circ\sim\nu$ be independent of $(T,U)$. If $F(T,J^\circ)$ has coverage at least $1-\alpha_0$ and expected diameter at most $R$ uniformly over the model, then
\[
\Pp\{K\notin F(T,J)\}\le\alpha_0+\Delta_{\rm seed},
\qquad
\E\diam F(T,J)\le R+D_n\Delta_{\rm seed}.
\]
In particular, choosing $\alpha_0\le\alpha-\Delta_{\rm seed}$ gives finite-sample coverage of at least $1-\alpha$.

There is also a correction at the original level. Suppose $J=(J_0,J_1)$ approximates $\operatorname{Unif}\{0,\ldots,L-1\}\otimes\nu$ with the same conditional error $\Delta_{\rm seed}$, and $F(T,J_1^\circ)$ is honest at level $1-\alpha$. Set
\[
a=\min\{L,\lceil L\Delta_{\rm seed}/\alpha\rceil\},\qquad \gamma=a/L.
\]
Return $\mathcal K_n$ when $J_0<a$, and $F(T,J_1)$ otherwise. This deterministic rule is honest at level $1-\alpha$ and satisfies
\[
\E\diam C\le\min\{D_n,\gamma D_n+(1-\gamma)R+D_n\Delta_{\rm seed}\}
\le\min\{D_n,R+D_n[(1+1/\alpha)\Delta_{\rm seed}+L^{-1}]\}.
\]
\end{lemma}
\begin{proof}
Conditional independence bounds the total variation distance between the joint laws of $(T,J)$ and $(T,J^\circ)$ by $\Delta_{\rm seed}$ in each state. Integration preserves this bound. Apply it to the noncoverage indicator and to diameter, which lies in $[0,D_n]$.

For the corrected rule, if $a=L$ the output is always $\mathcal K_n$. Otherwise $\gamma\ge\Delta_{\rm seed}/\alpha$. Under the ideal seed its noncoverage probability is at most $(1-\gamma)\alpha$, and seed replacement adds at most $\Delta_{\rm seed}$. Its ideal expected diameter is at most $\gamma D_n+(1-\gamma)R$. The diameter bound follows by the first part and $\gamma\le\Delta_{\rm seed}/\alpha+L^{-1}$.
\end{proof}
The bounds apply to the experiment based on $T$. The calibration tolerance and sampling radius below account for the removal of reference observations. Fixed linear maps for extracting randomness from independent biased bits are studied by Zhou and Bruck \cite{zhou2011linear}. We use a uniform conditional bound for the reserved reference responses.

Let $V=(V_1,\ldots,V_h)\in\{0,1\}^h$ contain predesignated reference responses. Conditional on a shared environment $U$, the coordinates are independent and
\[
\tau\le\Pp(V_j=1\mid U)\le1-\tau,\qquad 0<\tau<1/2.
\]
Put $\rho=1-2\tau$. Fix a binary $d\times h$ matrix $A$ before observing responses and define $J=AV$ over $\mathbb F_2$. For $u\in\mathbb F_2^d$, let $w_A(u)$ be the Hamming weight of $A^\top u$.

\begin{lemma}[Linear seed certificate]\label{supp-lem:linear}
For every allowed state,
\[
\TV\{\mathcal L(J\mid U),\operatorname{Unif}(\mathbb F_2^d)\}
\le\Delta_A:=\frac12\left\{\sum_{u\ne0}\rho^{2w_A(u)}\right\}^{1/2}.
\]
The seed is independent of all retained responses conditional on $U$. If $A$ has full row rank and its row space has minimum nonzero weight $w$, then
\[
\Delta_A\le\tfrac12\sqrt{2^d-1}\,\rho^w.
\]
\end{lemma}
\begin{proof}
For every $u\ne0$, conditional independence gives the Walsh coefficient
\[
\widehat P_U(u):=\E\{(-1)^{u^\top J}\mid U\}
=\prod_{j:(A^\top u)_j=1}(1-2p_j(U)).
\]
Thus $|\widehat P_U(u)|\le\rho^{w_A(u)}$. Orthogonality of the Walsh characters gives
\[
2^d\sum_{z\in\mathbb F_2^d}\{\Pp(J=z\mid U)-2^{-d}\}^2
=\sum_{u\ne0}\widehat P_U(u)^2.
\]
Cauchy--Schwarz bounds the total variation distance by half the square root of the right side. The minimum-weight bound follows by bounding each summand. Independence of the reserved and retained coordinates gives conditional independence of the seed and retained data.
\end{proof}

\begin{lemma}[A logarithmic reserve]\label{supp-lem:exist}
For $m\ge2$, put
\[
d=\lceil\log_2m\rceil,\qquad c_\tau=\frac{1+(1-2\tau)^2}{2}<1,
\qquad
h=\left\lceil\frac{\log\{2m^2(2^d-1)\}}{-\log c_\tau}\right\rceil.
\]
There is a fixed binary $d\times h$ matrix with $\Delta_A\le1/(2m)$. In particular, $h=O_\tau(\log m)$.
\end{lemma}
\begin{proof}
Temporarily draw the entries of $A$ independently as fair bits. For each fixed $u\ne0$, the coordinates of $A^\top u$ are independent fair bits, so
\[
\E_A\rho^{2w_A(u)}=c_\tau^h.
\]
Consequently
\[
\E_A\sum_{u\ne0}\rho^{2w_A(u)}
=(2^d-1)c_\tau^h\le\frac1{2m^2}.
\]
Markov's inequality shows that a draw has sum at most $m^{-2}$ with probability at least $1/2$. At least one such matrix therefore exists. Fix it as part of the observation protocol. The certified sum is less than one, so $A^\top u\ne0$ for every nonzero $u$ and the matrix has full row rank.
\end{proof}
Any matrix satisfying the certificate, including a shorter column prefix, may be used provided that both the matrix and its length are fixed before responses are observed.
\begin{lemma}[Block-parity benchmark]\label{supp-lem:parity}
For $m\ge2$, set
\[
d=\lceil\log_2m\rceil,\qquad
\ell=\left\lceil\frac{\log(dm)}{-\log(1-2\tau)}\right\rceil,
\qquad h_{\rm B}=d\ell.
\]
When $h_{\rm B}\le m$, partition a predesignated reserve of $h_{\rm B}$ reference responses into $d$ disjoint blocks of length $\ell$. Suppose all responses are independent conditional on $U$ and every reserved probability lies in $[\tau,1-\tau]$ in every state. Let $B_a$ be the parity of block $a$, and encode these bits as $J=\sum_{a=1}^d2^{a-1}B_a$. Then, uniformly over the allowed states,
\[
\TV\{\mathcal L(J\mid U),\operatorname{Unif}\{0,\ldots,2^d-1\}\}
\le \frac d2(1-2\tau)^\ell\le\frac1{2m}.
\]
The seed is independent of the retained responses conditional on $U$, and $h_{\rm B}=O_\tau((\log m)^2)$.
\end{lemma}
\begin{proof}
Writing $\rho=1-2\tau$, conditional independence gives $|\E\{(-1)^{B_a}\mid U\}|=|\prod_{j\text{ in block }a}(1-2p_j(U))|\le\rho^\ell$. Each parity is therefore within $\rho^\ell/2$ in total variation of a fair bit. The blocks are conditionally independent, so a product coupling bounds their joint distance by $d\rho^\ell/2$. The choice of $\ell$ gives the stated error, and $d,\ell=O_\tau(\log m)$ give the reserve order. Disjointness from the retained coordinates gives conditional independence of the seed and retained data.
\end{proof}
Block parity supplies the same seed accuracy using $O_\tau((\log m)^2)$ designated interior reference coordinates, compared with $O_\tau(\log m)$ in Lemma~\ref{supp-lem:exist}. The block rule uses the corrected threshold and retained-reference declarations in Section~\ref{supp-sec:retained_declarations}, with $h=h_{\rm B}$, when $h_{\rm B}\le m/2$ and $h_{\rm B}/m\le g_0/2$. At other sizes, it returns $\mathcal K_n$.

\subsection{A deterministic rule with finite-sample coverage}\label{supp-sec:retained_declarations}

Fix $\beta\in(0,1)$ and a binary $d\times h$ matrix $A$ before observing responses. Apply $A$ over $\mathbb F_2$ to the $h$ reserved responses $V$ and encode the output as $J=\sum_{a=1}^d2^{a-1}(AV)_a$. Set $h<m$ and $m_0=m-h$. Let $\widetilde r(U)$ and $r_D(U)$ be the retained and reserved reference means. Since
\[
r(U)=\frac{m_0}{m}\widetilde r(U)+\frac hm r_D(U),
\qquad
\widetilde r(U)-r(U)=\frac{h}{m-h}\{r(U)-r_D(U)\},
\]
the declarations $r(U)\in[r_-,r_+]$ and $r_D(U)\in[\tau,1-\tau]$ give
\[
\Delta_h^{\mathrm{ref}}=
\min\left\{\frac hm,\frac{h}{m-h}
\max\{r_+-\tau,1-\tau-r_-\}\right\},
\qquad |\widetilde r(U)-r(U)|\le\Delta_h^{\mathrm{ref}}.
\]
The first bound in the minimum follows from $\widetilde r-r=(h/m)(\widetilde r-r_D)$ and both means lying in $[0,1]$. Solving the same identity for $\widetilde r$ gives
\[
\ell_h=\max\left\{0,\frac{mr_--h(1-\tau)}{m-h}\right\},
\qquad
u_h=\min\left\{1,\frac{mr_+-h\tau}{m-h}\right\},
\qquad \widetilde r(U)\in[\ell_h,u_h].
\]
Put $\varepsilon_h=\min\{1,\varepsilon+\Delta_h^{\mathrm{ref}}\}$ and $g_h=\ell_h-\bar b$, and assume $g_h>0$. These quantities depend only on the declared bounds and reserve size. Let $X_0$ be the sum of the $m_0$ retained reference responses, and define
\[
J^0_k=[(n-k)b_L-k\varepsilon_h,(n-k)b_U+k\varepsilon_h],
\]
\[
H^0_\beta=\left\{k\in\mathcal K_n:
\dist\left(Y-kX_0/m_0,J^0_k\right)
\le\sqrt{\frac{n+k^2/m_0}{2}\log\frac2\beta}\right\}.
\]
Write $q_\beta=\eta+(1-\eta)\beta$ and $t=(1-\alpha/q_\beta)_+$. Using a certified bound $\Delta_{\rm seed}\ge\Delta_A$ from Lemma~\ref{supp-lem:linear}, form the predetermined integer
\[
a=\min\{2^d,\lceil2^d(t+\Delta_{\rm seed})\rceil\}.
\]
Return the full count range if $J<a$, and otherwise the integer hull of $H^0_\beta$. The protocol fixes the reserve and bit order, making this rule a deterministic function of the full unit-indexed response vector. Empty hulls are empty and have diameter zero.

On every original good state, the triangle inequality and $a_M,\widetilde r\in[0,1]$ give $|a_M(U)-\widetilde r(U)|\le\varepsilon_h$. Every state satisfies the retained-reference range. The retained inversion therefore fails conditionally with probability at most $\beta$ on good states and at most one on bad states. In particular its unconditional error is at most $q_\beta$.

For $f(U)=\Pp(J<a\mid U)$, Lemma~\ref{supp-lem:linear} and rounding imply
\begin{equation}\label{supp-eq:seedfallback}
t\le f(U)\le\bar f:=\min\{1,t+2\Delta_{\rm seed}+2^{-d}\}.
\end{equation}
If $a<2^d$, use $a/2^d-\Delta_{\rm seed}\ge t$ for the lower bound and $a/2^d+\Delta_{\rm seed}\le t+2\Delta_{\rm seed}+2^{-d}$ for the upper bound. If $a=2^d$, the lower bound is immediate and the rounding condition implies $t+2\Delta_{\rm seed}+2^{-d}\ge1$. Conditional independence of $J$ and the retained inversion then gives
\[
\Pp(K\notin C\mid U)\le(1-f(U))\Pp(K\notin H^0_\beta\mid U).
\]
Integrating over the original good and bad states yields
\begin{equation}\label{supp-eq:seedcoverage}
\Pp(K\notin C)\le(1-t)q_\beta\le\alpha.
\end{equation}
The $\Delta_{\rm seed}$ adjustment preserves coverage at each finite sample size.

The all-state diameter argument in the environmental proof applies with $m_0,\varepsilon_h,g_h$. Put
\[
B^0_\beta=\min\left\{D_n,
\frac{4n\varepsilon_h+2n\delta_b+4\sqrt{(n+n^2/m_0)\log(2/\beta)/2}}{g_h}
+2n e^{-m_0g_h^2/2}\right\}.
\]
Because $f(U)\le\bar f$ and $1-f(U)\le1-t$, the deterministic rule obeys
\begin{equation}\label{supp-eq:seedrisk}
\sup_{\mathcal P_{\eta,\tau,S}}\E\diam C
\le\min\{D_n,\bar fD_n+(1-t)B^0_\beta\}.
\end{equation}
The bound includes the diameter of every full-range outcome.

For $m\ge2$, use a certified matrix from Section~\ref{supp-sec:reserve_construction} and set $\beta=(s+u)/4$ and $\Delta_{\rm seed}=1/(2m)$. Lemma~\ref{supp-lem:linear} gives $\Delta_A\le1/(2m)$ and $h=O_\tau(\log m)$. Since $2^{-d}\le1/m$, we have $\bar f\le t+2/m$. Use the preceding rule when $h\le m/2$ and $h/m\le g_0/2$. At other sizes, return $\mathcal K_n$. These conditions and all thresholds are predetermined.

Since $\ell_h\ge r_--h/m$, these conditions imply $m_0\ge m/2$ and $g_h\ge g_0/2$. They hold for every sufficiently large $m$. As in the environmental upper bound, $t\le(v+s/4)/\alpha$. The additional calibration cost satisfies
\[
n\Delta_h^{\mathrm{ref}}\le\frac{nh}{m}
=O_\tau(n\log m/m)=O_\tau(n/\sqrt m),
\]
and the extra fallback cost is at most $2D_n/m=O(n/\sqrt m)$. The retained sampling radius has the same order as the original one, and $\log(2/\beta)\le C\log[\mathrm e/\min\{1,s+u\}]$. Substitution in~\eqref{supp-eq:seedrisk} proves the upper rate uniformly over $\eta$. The exceptional reference sizes are bounded by a constant depending only on $\tau,g_0$. On this finite set, $s$ is bounded below by a positive constant, so the full-range rule also has the stated rate up to a fixed constant. This includes $m=1$.

\subsection{Lower bounds with reference randomness}\label{supp-sec:reserve_lower}

Set $b=b_L\in[0,\bar b]$ and $r_*=(r_-+r_+)/2\in(\tau,1-\tau)$. We check each family used in Theorem~\ref{thm:environment}.

\begin{enumerate}[leftmargin=*,itemsep=2pt]
\item The common bad law has all main probabilities $b$ and all reference probabilities $r_*$. Its mismatch is $r_*-b>\bar\varepsilon$. Every reserved reference probability is interior, including when $b=0$.
\item In the main-response family, all reference probabilities equal $r_*$. Its varying main probability lies between $r_*$ and $r_*+(1-r_*)/16<1$, and all other main probabilities equal $b$. The full-law divergence bound $C(j-k)^2/n$ therefore holds.
\item In the reference-response family, the reference probability $r(k)$ lies in $(r_-,r_*]\subset(\tau,1-\tau)$. Main probabilities are $r_*$ or $b$. The full-law divergence bound $Cm(j-k)^2/n^2$ holds.
\item The identical-law calibration and background pairs have reference probability $r_*$ throughout. Their admissible membership interpretations and the bounds $n\varepsilon$ and $n\delta_b$ are unchanged.
\end{enumerate}
These families use the same designated set $S_m$ in every model and have interior probabilities at all reference coordinates. Applying the two-point bound~\eqref{supp-eq:two} to the main-response family with count separation $\lfloor c\sqrt n\rfloor$, and, for $m\le n$, to the reference-response family with separation $\lfloor cn/\sqrt m\rfloor$, gives the two baseline sampling terms within both reference-interior classes, for sufficiently small fixed $c>0$. The divergence bounds and Pinsker's inequality control the full observation laws. For $m>n$, the reference term is dominated by $\sqrt n$. Together with the identical-law ambiguity pairs, these constructions establish~\eqref{supp-env:eq:baseline} within both classes. The remaining many-count and small-error testing arguments in Section~\ref{supp-env:sec:lower} give the lower rate. Main coordinates with probability zero have the same law in each compared model, and the varying coordinates have fixed interior margins.

The linear construction uses only $S_m$ and proves the upper bound for the larger subset-reference class. Its restriction to the all-reference class gives the same bound. This proves Theorem~\ref{thm:interior}.\hfill$\square$

The subset and its ordering are specified before observing responses. The number of required interior reference coordinates is $h_m=O_\tau(\log m)$. The background may equal zero or lie below $\tau$.

\subsection{Reliability declaration and misspecification}\label{supp-sec:sensitivity}

Let $\lambda=\Pp\{|a_M(U)-r(U)|>\varepsilon\}$ be the actual calibration-failure probability and let $\eta\in[0, 1]$ be its declared upper bound. Fix $0<\beta=\beta(\eta)<1$ before observing the responses. The all-state reference and background restrictions remain in force. Put
\[
q_\beta(x)=x+(1-x)\beta,\qquad
t_{\beta,\eta}=\left(1-\frac{\alpha}{q_\beta(\eta)}\right)_+.
\]
For external full-range correction, or for a corrected internal seed satisfying the conditions below,
\begin{equation}\label{supp-eq:misspec}
\Pp(K\notin C)\le(1-t_{\beta,\eta})q_\beta(\lambda).
\end{equation}
Thus $\lambda\le\eta$ guarantees coverage at least $1-\alpha$. If $q_\beta(\eta)>\alpha$, the noncoverage bound is
\[
\alpha\frac{q_\beta(\lambda)}{q_\beta(\eta)}.
\]
For direct inversion with $\eta<\alpha$, take $\beta=(\alpha-\eta)/(1-\eta)$ and $t=0$ in~\eqref{supp-eq:misspec}.

\paragraph{Proof.} Let $e(U)$ be the conditional noncoverage probability of the base set. H and its Chernoff/H refinement satisfy $e(U)\le\beta$ on the good event and $e(U)\le1$ elsewhere. External correction therefore gives
\[
\E[(1-t_{\beta,\eta})e(U)]
\le(1-t_{\beta,\eta})\{(1-\lambda)\beta+\lambda\}.
\]
For the internal rule, the reserved-response probabilities obey their interior bounds in every state. The seed remains independent of retained responses conditional on $U$. Its corrected threshold ensures conditional full-range probability $f(U)\ge t_{\beta,\eta}$. Consequently its error is $\E[(1-f(U))e(U)]$, which obeys the same bound. Ineligible sizes return the full range and have zero error.

For Chernoff/H inversion, the retained-reference declarations from Section~\ref{supp-sec:retained_declarations} are
\begin{equation}\label{supp-eq:retained_range}
\begin{gathered}
m_0=m-h,\qquad
\varepsilon_h=\min\{1,\varepsilon+\Delta_h^{\mathrm{ref}}\},\\
\widetilde r\in[\ell_h,u_h],\qquad g_h=\ell_h-\bar b.
\end{gathered}
\end{equation}
The bounds $\Delta_h^{\mathrm{ref}},\ell_h,u_h$ depend only on $r_-,r_+,\tau,h,m$. These declarations contain every retained good-state law. The conditional error bound $e(U)\le\beta$ follows from the statewise construction.

The declaration specifies a model class by bounding the probability of calibration failure across shared states. A single roster does not identify this probability. Comparing declared bounds around $\alpha$ quantifies their effect on width and coverage. Theorem~\ref{thm:environment} gives risk of order $n$ for a fixed overshoot $\eta-\alpha=c>0$, and a shrinking overshoot contributes $n(\eta-\alpha)_+$. The bound in~\eqref{supp-eq:misspec} is a coverage guarantee, while the actual coverage also depends on the conditional response laws.

\subsection{Roster subexperiment on the proportion grid}\label{supp-sec:roster-grid-detail}

Retain the fixed margins of Theorem~\ref{supp-thm:rate}, with $m\asymp n$, $\varepsilon=\delta_b=0$, known background $b$ and $\eta=\alpha$. Fix $\rho_-<\rho_0<\rho_+$, choose $r_0\in(r_-,r_+)$, and set $k_*=\lceil\rho_0n\rceil$. Choose a fixed $\ell>0$ satisfying
\[
\rho_0+\ell<\rho_+,\qquad
\frac{\ell}{\rho_0}(r_0-b)<\frac{1-r_0}{2}.
\]
For sufficiently large $n$, the count family
\[
\mathcal I_n=\{k_*,\ldots,k_*+\lfloor\ell n\rfloor\}
\]
lies in $\mathcal K_n$. For each $k\in\mathcal I_n$, let the first $k$ main units be the fixed members. Define the clean product law $G_{k,n}$ by
\[
q_i(k)=
\begin{cases}
b+\dfrac{k}{k_*}(r_0-b),&i\le k_*,\\
b,&i>k_*,
\end{cases}
\qquad p_j(k)=r_0\quad(1\le j\le m).
\]
The sum of the member probabilities is $k_*\{b+(k/k_*)(r_0-b)\}+(k-k_*)b=kr_0$. Thus the member and reference averages equal $r_0$, and the nonmember average equals $b$. The changing probabilities satisfy
\[
r_0\le q_i(k)
\le r_0+\frac{\ell}{\rho_0}(r_0-b)<\frac{1+r_0}{2}<1
\qquad(i\le k_*),
\]
so their interior margin is uniform in $n$ and $k$.

Let $Q_n$ be the product law with all main probabilities equal to $b$ and all reference probabilities equal to $r_0$. Under each membership vector above, it satisfies the reference and background restrictions in~\eqref{eq:envclass}, while member--reference calibration fails. Consequently $(1-\alpha)G_{k,n}+\alpha Q_n$ is an admissible two-state model with membership fixed across states. The pure clean law $G_{k,n}$ is also admissible.

Only the first $k_*$ coordinates change between $G_{k,n}$ and $G_{j,n}$. Using $\operatorname{kl}(p,q)\le(p-q)^2/\{q(1-q)\}$ and the uniform interior margin gives
\[
\KL(G_{k,n}\Vert G_{j,n})
\le Ck_*\left\{\frac{(j-k)(r_0-b)}{k_*}\right\}^2
\le C\frac{(j-k)^2}{n}.
\]
Coordinates whose probability is $b=0$ have the same law throughout the family.

Take the fixed proportion intervals
\[
I=[\rho_0+\ell/4,\rho_0+3\ell/4],\qquad
J=[\rho_0+\ell/4,\rho_0+\ell/2],\qquad h_0=\ell/4.
\]
Then $J+[0,h_0]\subseteq I$, and $I\cap n^{-1}\mathbb Z$ is contained in $\{k/n:k\in\mathcal I_n\}$ for all sufficiently large $n$. Every shift $h\in n^{-1}\mathbb N\cap(0,h_0]$ from $J\cap n^{-1}\mathbb Z$ therefore gives an actual pair of parameters, with KL at most $Cnh^2$.

For an honest confidence set on this subexperiment, write $R$ for its worst-case expected diameter on the proportion scale. The mixture risk gives $\alpha\E_{Q_n}\diam C\le R$. By Lemma~\ref{lem:gp-budget}, its clean errors satisfy
\[
\frac1n\sum_{\theta\in I\cap n^{-1}\mathbb Z}e_\theta
\le\frac{R}{1-\alpha}+\frac{\alpha}{(1-\alpha)n}.
\]
Since $1/n=o(\sqrt{\log n/n})$, the interval, mesh and one-sided KL conditions of Corollary~\ref{cor:gp-rate} hold with $N=n$ and $\kappa=2$. Hence $R\gtrsim\sqrt{\log n/n}$. Given an honest count set on the full roster class, divide its elements by $n$ and intersect with $I$. Coverage is preserved for the subexperiment, and expected diameter is at most the original risk divided by $n$. This proves~\eqref{eq:roster-critical-lower}.

For the upper bound, set $\varepsilon=\delta_b=0$ in \eqref{supp-env:eq:Hupper}, use $m\asymp n$, and divide by $n$. The resulting all-state bound is
\begin{equation}\label{supp-eq:roster-all-state-width}
\sup_{\text{allowed states}}\E[\diam(H_\beta/n)]
\le\min\{D_n/n,\ C\sqrt{\log(2/\beta)/n}+Ce^{-cn}\}.
\end{equation}
The reference restriction holds in every state, so integrating this conditional expectation bound controls both components of the failure mixture. Proposition~\ref{prop:upper-transfer} with $\beta=n^{-1/2}$ supplies the matching upper rate.

\section{Computation}\label{supp-sec:chernoff}

The rate proof uses Hoeffding inversion because its diameter is explicit. A weighted Chernoff construction can improve width while preserving the same calibration class. Let $\theta\ne0$. Concavity of the Bernoulli log-MGF in its success probability, applied within each group, gives
\begin{align}\label{supp-eq:mgf}
G_k(\theta;r,a,b)={}&k\log\{1+a(e^\theta-1)\}+(n-k)\log\{1+b(e^\theta-1)\}\notag\\
&+m\log\{1+r(e^{-\theta k/m}-1)\}.
\end{align}
The negative reference weight enters through $e^{-\theta k/m}$. Let
\[
S_k(\theta)=\sup_{r,a,b:\ |a-r|\le\varepsilon,\ b\in[b_L,b_U]}G_k(\theta;r,a,b),
\]
with probabilities in $[0, 1]$ and any additional declared reference range. For fixed finite positive and negative tilt grids $\Theta_+,\Theta_-$, define
\[
t_k^+=\min_{\theta\in \Theta_+}\frac{S_k(\theta)+\log(2/\alpha)}{\theta},\qquad
t_k^-=\max_{\theta\in \Theta_-}\frac{S_k(\theta)+\log(2/\alpha)}{\theta}.
\]
Conditional Markov bounds justify accepting $k$ when $t_k^-\le W_k\le t_k^+$. Each one-sided threshold controls $\alpha/2$. A finite grid only loses sharpness relative to optimisation over all tilts.

For each candidate and tail, taking the stricter of this threshold and the Hoeffding threshold chooses a deterministic bound before seeing the data. Thus the Chernoff/H intersection remains honest at level $1-\alpha$ and is contained in $C_H$. Coverage plus containment implies that the refinement attains the rate in Theorem~\ref{supp-thm:rate}.

For $\theta>0$, optimise $a=\min(1,r+\varepsilon)$ and $b=b_U$. For $\theta<0$, optimise $a=\max(0,r-\varepsilon)$ and $b=b_L$. The resulting function of $r$ is concave with at most two pieces. On each piece $[l,u]$, a tangent at $x$ gives the upper bound
\[
\sup_{r\in[l,u]}G_k(\theta;r)\le G_k(\theta;x)
+\max\{G_k'(\theta;x)(l-x),G_k'(\theta;x)(u-x)\}.
\]
This gives an outer bound for the nuisance supremum.

Let $A_\beta^{\mathrm{Ch}}=C_{\mathrm{Ch},\beta}\cap\mathcal K_n\subseteq H_\beta$. On every good state, $A_\beta^{\mathrm{Ch}}$ has conditional noncoverage at most $\beta$. The external correction therefore gives error at most $(1-t_\beta)q_\beta$. For the internal rule, compute $A_\beta^{\mathrm{Ch}}$ with $m_0=m-h$, $\varepsilon_h$ and $[\ell_h,u_h]$ from~\eqref{supp-eq:retained_range}. Conditional independence of the seed and retained data gives the same bound with the corrected full-range threshold. The misspecification bound~\eqref{supp-eq:misspec} follows under the conditions of Section~\ref{supp-sec:sensitivity}. Coupling the H and Chernoff/H rules with the same seed makes the latter's integer hull a subset of the former's on every outcome. Expected diameter can only decrease, and the lower bound in Theorem~\ref{thm:interior} still applies. Thus Chernoff substitution preserves the stated minimax rate.

\subsection{Tilt grid and nuisance maximisation}

For candidate $k$, the positive tilts are
\[
\Theta_+(k)=\{10^{-5}(20/10^{-5})^{j/159}:j=0,\ldots,159\}/\max(1,k/m),
\qquad \Theta_-(k)=-\Theta_+(k).
\]
Thus $|\theta|\le20$ and $|\theta k/m|\le20$. The grid is fixed before observing $X,Y$. Minimising deterministic Markov thresholds selects one valid threshold for each tail.

After optimising the endpoints for $a,b$, put $t=e^\theta-1$ and $s=e^{-\theta k/m}-1$. On the segment where $a=r+\operatorname{sign}(\theta)\varepsilon$ is affine, the stationary candidate is
\[
r_*=-\frac{kt+ms+mst\operatorname{sign}(\theta)\varepsilon}{(k+m)ts}.
\]
It is clipped to that segment. The derivative and tangent bound at the clipped point give an upper envelope for the segment supremum. On the constant-$a$ segment, the reference term is monotone and its maximum is at the endpoint nearest the kink. The larger bound is used. At $k=0$, only the nonmember term remains.

Numerical evaluation is specified in Section~\ref{supp-sec:numerical_arithmetic}.

\subsection{Fast H inversion}

The finite-sample reliability comparison uses a singleton background $b$. Put $z=Y-nb$, $\widehat g=X/m-b$ and $c=\log(2/\beta)/2$. The endpoint inequalities are
\[
z-k(\widehat g+\varepsilon)-\sqrt{c(n+k^2/m)}\le0,\qquad
-z+k(\widehat g-\varepsilon)-\sqrt{c(n+k^2/m)}\le0.
\]
If $\widehat g-\varepsilon>\sqrt{c/m}$, the first left side decreases and the second increases with $k$. Two integer binary searches therefore find the H endpoints. For each observation that fails this sufficient condition, the procedure evaluates both inequalities over the full candidate range and returns the hull of the accepted counts. Empty sets are represented by a lower endpoint exceeding the upper endpoint.

\subsection{Cached inversion}

For each design and error allocation, cache the lower and upper Chernoff/H thresholds $L_k,U_k$. Each sign uses $G=160$ tilts. Nuisance maximisation at one tilt requires a fixed number of evaluations. Setup for one allocation costs $O(|\mathcal K_n|G)$, and multiple allocations reuse the same log-MGF values. After reserving references, setup uses $m_0$, $\varepsilon_h$ and $[\ell_h,u_h]$ from~\eqref{supp-eq:retained_range}.

For observed $x$, if both $L_{k+1}-L_k+x/m_0$ and $U_{k+1}-U_k+x/m_0$ are nonnegative throughout the candidate range, the acceptance inequalities have monotone endpoints. Two binary searches then compute the integer hull in $O(\log n)$ time. The minimum slopes are cached with the thresholds. Otherwise, we scan the candidates in H and take the hull, at cost $O(\log n+|H_\beta|)$ when the H monotonicity condition holds. Without caching, evaluating the tilt grid on that H set costs $O(\log n+|H_\beta|G)$. Full scanning applies when that condition fails.

\section{Numerical results}\label{supp-sec:numerical-results}

The roster simulations estimate coverage and expected diameter under specified response laws. We compare external randomisation and linear extraction in a two-state model and at a common clean law.

\subsection{Evaluation of the analytic bounds}\label{supp-sec:analytic-evaluation}

Main Figures~\ref{fig:sharp-bounds}--\ref{fig:slack-frontier} evaluate the finite bounds without simulation, with $\alpha=0.05$ and Gaussian parameters $D=\sigma=1$. Each curve uses 85 log-spaced sample sizes rounded to distinct integers, ranging over $10^2$--$10^{14}$ for Figure~\ref{fig:sharp-bounds}, and $10^2$--$10^{12}$ or $10$--$10^5$ for the polynomial or exponential sequences in Figure~\ref{fig:slack-frontier}.

Each one-dimensional search evaluates 75 equally spaced arguments, then refines the interval adjacent to the best grid point by bounded scalar minimisation with absolute tolerance $10^{-11}$, retaining the better value. For the Uniform lower bound~\eqref{supp-ep:finitelower}, the search varies $x=Nh/a$ and uses the analytic maximising $a$ at fixed $x$. For the Gaussian lower bound~\eqref{supp-gaussian:finitelower}, it varies $\sqrt N d/\sigma$ and equates the two bound terms by a bracketed root search in $\log\varepsilon$ (tolerance $10^{-12}$). Both upper curves select $\log\beta$ to minimise the full corrected diameter bound. The frontier lower curves select $s\in(0,t)$ in \eqref{slack:finite}, and their upper curves are $D_N(t)$. Polynomial sequences use $t=\gamma\log N$, $\gamma\in\{0.5,1,2\}$. Exponential sequences use $t=cN$, $c\in\{0.15,0.5\}$. The curves show admissible finite bounds at the selected parameters.

\subsection{Numerical evaluation}\label{supp-sec:numerical_arithmetic}

The tables use double-precision evaluation of the Chernoff thresholds. Bernoulli log-MGFs are computed as $\operatorname{logaddexp}(\log(1-p),\log(p)+\theta)$, including $p=0,1$, and the nuisance envelope receives padding $10^{-10}(1+n+m+|\widehat S_k(\theta)|)$, where $\widehat S_k(\theta)$ is the computed envelope before padding. The reported padded double-precision values approximate the mathematical outer envelope used in the coverage theorems. Inversion uses binary search where monotonicity holds and full candidate scans otherwise.

\subsection{Reliability and interior procedures}\label{supp-sec:finitesample}

We set $\alpha=0.05$, $\tau=0.1$, $b_L=b_U=\bar b=0.1$, $[r_-,r_+]=[0.45, 0.55]$, $\varepsilon=0$, and $\mathcal K_n=\{0.2n,\ldots,0.7n\}$. The first $K=0.4n$ units are members. A shared state $U\in\{0,1\}$ has $\Pp(U=1)=\lambda=\eta$ in the displayed reliability grid. Conditional on $U$, all responses are independent.

In state 0, member probabilities alternate between 0.3 and 0.7, nonmember probabilities equal 0.1, and reference probabilities alternate between 0.3 and 0.7. In state 1, all main probabilities equal 0.1 and reference probabilities alternate between 0.1 and 0.9. Thus $r(U)=0.5$ in both states, $a_M(0)=0.5$, $a_M(1)=0.1$, and $b_M(U)=0.1$. Every individual probability lies in $[0.1, 0.9]$. In the bad state, the reference probabilities attain the endpoints of $[\tau,1-\tau]$, so $|1-2p_j|=1-2\tau=0.8$. The extraction step is therefore evaluated at the largest coordinate-wise bias allowed by the reference condition.

At and above the boundary, the full-range procedures use $\beta=(s+u)/4$, with $s$ computed from the original $n,m$. External correction follows Section~\ref{supp-env:sec:upper}, and the linear rule uses the threshold in Section~\ref{supp-sec:retained_declarations} with $\Delta_{\rm seed}=1/(2m)$. The matrices in Section~\ref{supp-sec:certificates} and reserve positions at the start of the reference roster are fixed before responses. Linear extraction uses~\eqref{supp-eq:retained_range}, computed from $r_-,r_+,\tau,h,m$. If $h>m/2$ or $h/m>g_0/2$, the internal rule returns $\mathcal K_n$.

For $\eta<\alpha$, direct inversion uses all references and $\beta_{\rm direct}=(\alpha-\eta)/(1-\eta)$, giving error at most $\alpha$. Each corrected procedure chooses among this rule and its corrections at 161 logarithmically spaced values of $\beta$ in $[10^{-6},0.5]$, augmented by $\beta_{\rm direct}$ and $(s+u)/4$. Selection minimises a bound determined by the declarations before responses are observed, with ties favouring direct inversion. The displayed reliability grid therefore changes both the declared failure bound and the selection rule at the boundary.

For this known-background design, let $g=r_--b$, $k_+=\max\mathcal K_n$ and $\mathcal Z_g=\{jg/200:1\le j\le199\}$. The conditional base-diameter bound used for selection is
\[
B^{\rm sel}_\beta=\min\left\{D_n,\min_{z\in\mathcal Z_g}
\left[\min\left\{D_n,\frac{2k_+\varepsilon+2R_{k_+}(\beta)}{z}\right\}
+D_n e^{-2m(g-z)^2}\right]\right\}.
\]
Indeed, on $X/m\ge b+z$ the pairwise residual argument in~\eqref{supp-env:eq:residual} bounds diameter by $(2k_+\varepsilon+2R_{k_+}(\beta))/z$. The complementary event has conditional probability at most $e^{-2m(g-z)^2}$ in every state. We minimise $t_\beta D_n+(1-t_\beta)B^{\rm sel}_\beta$ for the external rule. For the reserve rule, replace $(m,\varepsilon,g)$ by $(m_0,\varepsilon_h,g_h)$ and use the correction bound~\eqref{supp-eq:seedrisk}. Both bounds apply to the contained Chernoff/H refinement. Selecting direct inversion uses no reserve and no correction coin.

At $\eta=\alpha$, the rate construction has $\beta=s/4=1/(2\sqrt n)$ when $m=n$, approximately $0.00790$ at $n=4{,}000$ and $0.00198$ at $64{,}000$. Its clean-state error therefore tends to zero, explaining the high clean coverage, while the uniform guarantee controls the mixture of clean and failed states.

For each $n=m\in\{1{,}000,4{,}000,16{,}000,64{,}000\}$, we generate 50,000 sets of response totals per state from independent binomial probability blocks. Common random numbers are used across procedures and values of $\eta$ to reduce Monte Carlo variation in the comparisons. The declarations are
\[
\eta\in\{0,0.025,0.04,0.045,0.05,0.055,0.06,0.075,0.10\}.
\]

The reference pattern permits exact calculation of the conditional seed laws. For the linear map, let $N_A(u)$ count the high-probability positions selected by $A^\top u$. If the lower reference probability is $p$, then
\[
\widehat P_U(u)=(-1)^{N_A(u)}(1-2p)^{w(A^\top u)},\qquad
\Pp(J=z\mid U)=2^{-d}\sum_{u\in\mathbb F_2^d}(-1)^{u^\top z}\widehat P_U(u).
\]
We evaluate this finite sum by a Walsh transform. This law determines $f_U=\Pp(J<a\mid U)$ at the prescribed full-range threshold. For retained-data diameter $W_U$ and coverage indicator $I_U$, we use
\[
f_UD_n+(1-f_U)W_U,\qquad f_U+(1-f_U)I_U.
\]
For the randomised procedure, replace $f_U$ by $t_\beta$. For direct inversion, use $f_U=0$ and its own $\beta$. We then average using the state weights $1-\eta,\eta$. This integrates the environment and full-range decisions while retaining Monte Carlo variation from the responses. If the two conditional-state estimates have standard errors $s_0,s_1$, the combined standard error is $\{(1-\eta)^2s_0^2+\eta^2s_1^2\}^{1/2}$.

Write $C_R^{\mathrm{Ch}}$ and $C_D^{\mathrm{Ch}}$ for the external and linear procedures after Chernoff substitution. The tables abbreviate them as R and D. Hoeffding inversion $C_H$ and its Chernoff/H refinement use the same response and seed-generating designs, with retained-reference declarations~\eqref{supp-eq:retained_range}. At $\eta=\alpha$, replacing H by Chernoff/H reduces the normalised mean diameters for the external and linear procedures from $0.3648,0.4151$ to $0.3225,0.3732$ at $n=m=4{,}000$, and from $0.1087,0.1131$ to $0.0955,0.0999$ at $64{,}000$. The external--linear differences are $0.0507D_n$ and $0.0044D_n$, respectively (Table~\ref{supp-tab:critical_chernoff}). The subcritical selection chooses direct inversion for both procedures at all displayed sizes $n\ge4{,}000$. At $n=1{,}000$, the external rule selects positive correction at $\eta=0.04,0.045$, because these choices minimise the declared-parameter bound.
\begin{table}[H]\centering\small
\begin{tabular}{rrrrr}\toprule
$n=m$ & Linear $h/m$ & R diameter & D diameter & Min. coverage\\\midrule
1,000 & 0.1050 & 0.5518 & 0.7355 & 0.9606\\
4,000 & 0.0315 & 0.3225 & 0.3732 & 0.9562\\
16,000 & 0.0091 & 0.1785 & 0.1934 & 0.9534\\
64,000 & 0.0026 & 0.0955 & 0.0999 & 0.9518\\
\bottomrule\end{tabular}
\caption{Chernoff/H results at $\eta=\alpha$, under the same simulation design and reserve specification as the H comparison. R and D denote external randomisation and linear extraction. Diameter columns report simulated means divided by $D_n$, and minimum coverage is the smaller estimate across R and D. The $h/m$ column reports the linear reserve fraction.}\label{supp-tab:critical_chernoff}
\end{table}
For the four linear reserves, $\Delta_h^{\mathrm{ref}}$ is $0.052793$, $0.014636$, $0.004144$ and $0.001184$. These are the calibration adjustments, and $h/m$ is the reserve fraction (Table~\ref{supp-tab:critical_chernoff}). At $m=1{,}000$, $h=105$ gives $\varepsilon_h=0.052793$ and $n\varepsilon_h=52.79$, against $D_n=500$. The certified seed TV error is below $0.0005$, and correction plus rounding adds at most $(1/m+2^{-d})D_n<0.002D_n$ beyond the external full-range probability. The reserve adjustment is the larger finite-sample cost at this size.

The linear rule and block-parity seed of Lemma~\ref{supp-lem:parity} use the same declarations and error allocations (Table~\ref{supp-tab:block_parity}). Each method uses its own retained-reference adjustment. Both procedures are evaluated under the same simulated design using Chernoff/H inversion.
\begin{table}[H]\centering\small
\begin{tabular}{rrrrr}\toprule
$n=m$ & $h_{\rm L}$ & $h_{\rm B}$ & Linear & Block parity\\\midrule
1,000 & 105 & 420 & 0.7355 & 1.0000\\
4,000 & 126 & 588 & 0.3732 & 0.6129\\
16,000 & 146 & 784 & 0.1934 & 0.2625\\
64,000 & 168 & 1,008 & 0.0999 & 0.1221\\
\bottomrule\end{tabular}
\caption{Planned reserve lengths and mean diameters divided by $D_n$
at $\eta=\alpha=0.05$. Here $h_{\rm L}$ and $h_{\rm B}$ denote the
linear-extractor and block-parity reserve sizes, respectively.
Both rules use the two-state design above and
Chernoff/H inversion. At $m=1{,}000$, the block reserve fails the
eligibility condition, so the rule returns the full range.}
\label{supp-tab:block_parity}
\end{table}

Across the selected reliability designs, the smallest coverage estimate is $0.9515$ over the displayed methods, with direct inversion included for $\eta<\alpha$ (Table~\ref{supp-tab:selected_grid}).
\begin{table}[H]\centering\small
\begin{tabular}{rrrrrr}\toprule
$n=m$ & $\eta$ & $C_R^{\mathrm{Ch}}/D_n$ & $C_D^{\mathrm{Ch}}/D_n$ & Direct $C_{\mathrm{Ch},\beta}/D_n$ & Min. coverage\\\midrule
1,000 & 0.000 & 0.3817 & 0.3817 & 0.3817 & 0.9957\\
1,000 & 0.045 & 0.5683 & 0.4666 & 0.4666 & 0.9547\\
1,000 & 0.050 & 0.5518 & 0.7355 & --- & 0.9606\\
1,000 & 0.060 & 0.6078 & 0.7655 & --- & 0.9591\\
4,000 & 0.000 & 0.1893 & 0.1893 & 0.1893 & 0.9964\\
4,000 & 0.045 & 0.2301 & 0.2301 & 0.2301 & 0.9548\\
4,000 & 0.050 & 0.3225 & 0.3732 & --- & 0.9562\\
4,000 & 0.060 & 0.4205 & 0.4633 & --- & 0.9552\\
16,000 & 0.000 & 0.0946 & 0.0946 & 0.0946 & 0.9956\\
16,000 & 0.045 & 0.1147 & 0.1147 & 0.1147 & 0.9548\\
16,000 & 0.050 & 0.1785 & 0.1934 & --- & 0.9534\\
16,000 & 0.060 & 0.3060 & 0.3184 & --- & 0.9528\\
64,000 & 0.000 & 0.0473 & 0.0473 & 0.0473 & 0.9964\\
64,000 & 0.045 & 0.0573 & 0.0573 & 0.0573 & 0.9548\\
64,000 & 0.050 & 0.0955 & 0.0999 & --- & 0.9518\\
64,000 & 0.060 & 0.2409 & 0.2445 & --- & 0.9515\\
\bottomrule\end{tabular}
\caption{Selected reliability designs. Diameter columns report simulated means divided by $D_n$. A dash indicates that direct inversion is used only for $\eta<\alpha$. For $\eta<\alpha$, R and D use the bound-based selection, which includes direct inversion. For $\eta\ge\alpha$, they use the rate construction.}\label{supp-tab:selected_grid}
\end{table}

At $n=m=4{,}000$, we construct each rule using $\eta$ and then reweight its two conditional-state results using $\lambda$. This changes the environmental mass while keeping membership, response probabilities and each declared rule fixed. The selected declarations come from the $5\times5$ grid on $\{0.04,0.045,0.05,0.055,0.06\}$ for both R and D (Table~\ref{supp-tab:eta_sensitivity}). In this grid, the two rules use the same $\beta$, so the coverage lower bound $1-(1-t_{\beta,\eta})q_\beta(\lambda)$ applies to both. Their full-range probabilities may differ.
\begin{table}[H]\centering\small
\begin{tabular}{rrrrrr}\toprule
Declared $\eta$ & Actual $\lambda$ & Coverage bound & R coverage & D coverage & R diameter$/D_n$\\\midrule
0.04 & 0.04 & 0.9500 & 0.9595 & 0.9595 & 0.2173\\
0.04 & 0.05 & 0.9401 & 0.9495 & 0.9495 & 0.2150\\
0.04 & 0.06 & 0.9302 & 0.9395 & 0.9395 & 0.2128\\
0.05 & 0.04 & 0.9586 & 0.9649 & 0.9652 & 0.3245\\
0.05 & 0.05 & 0.9500 & 0.9562 & 0.9565 & 0.3225\\
0.05 & 0.06 & 0.9414 & 0.9475 & 0.9479 & 0.3205\\
0.06 & 0.04 & 0.9647 & 0.9701 & 0.9704 & 0.4239\\
0.06 & 0.05 & 0.9574 & 0.9626 & 0.9629 & 0.4222\\
0.06 & 0.06 & 0.9500 & 0.9552 & 0.9555 & 0.4205\\
\bottomrule\end{tabular}
\caption{Reliability sensitivity for Chernoff/H at $n=m=4{,}000$. Declared $\eta$ is the failure-probability upper bound used to construct each rule, and actual $\lambda$ is the failure probability used to weight the two states. R and D coverage columns are simulation estimates. The bound is the finite-sample coverage lower bound obtained from~\eqref{supp-eq:misspec}. Diameter estimates include full-range decisions. Coverage MCSE is at most 0.00011.}\label{supp-tab:eta_sensitivity}
\end{table}
Declaring $0.04$ when the actual failure probability is $0.05$ gives R and D coverage 0.9495. Declaring $0.05$ gives R coverage 0.9562. At actual failure probability $0.05$, increasing the declaration from $0.05$ to $0.06$ increases R mean diameter from $0.3225D_n$ to $0.4222D_n$. Understating the failure probability lowers coverage in this design. Equation~\eqref{supp-eq:misspec} quantifies the coverage guarantee under misspecification. With $\eta=0$, $\beta=\alpha=0.05$ and $\lambda=0.5$, its coverage lower bound is $0.475$.

\subsection{Dirac-state lower bound and comparison at a common law}\label{supp-sec:samelaw}

Use $\eta=\alpha=0.05$, $b=0.1=1/10$, $[r_-,r_+]=[0.45, 0.55]$ and $\mathcal K_n=\{0.2n,\ldots,0.7n\}$. Then $r_0=c_r=0.5$, $d_r=0.05$, $d_b=10$, $k_-=0.2n$ and $D_n=0.5n$. For $n=m$ at the sample sizes below, Proposition~\ref{supp-det:finite} gives
\[
\frac{\mathrm{LB}}{D_n}=
\left(1-\frac{40}{n}\right)
\left[1-2e^{-0.005n}-2e^{-0.02n}\right]_+.
\]
Evaluate $C_R^{\mathrm{Ch}}$ and $C_D^{\mathrm{Ch}}$ at the same clean law $G_-$, with $K=k_-$, member and reference probabilities $0.5$, and nonmember probabilities $0.1$. Both rules retain the declaration $\eta=0.05$, although the generating law has actual failure probability zero. Both use $\beta=1/(2\sqrt n)$, and the internal rule uses the fixed matrices in Section~\ref{supp-sec:certificates} and all retained-reference adjustments. Fair independent reserved responses give an exactly uniform seed because each matrix has full row rank. The corrected threshold is used, and its full-range decision is integrated exactly.

At each size, the rules share $50{,}000$ simulated main and retained-reference totals. For the external rule, independent $\operatorname{Bin}(h,0.5)$ reserve totals complete the full reference sample, and its independent full-range branch has probability $t_\beta$. The MCSE of mean diameter divided by $D_n$ is at most $0.000230$ for R and $0.000283$ for D. Coverage MCSE is at most $0.000138$.
\begin{table}[H]\centering\small
\begin{tabular}{rrrrrr}\toprule
$n=m$ & Deterministic LB$/D_n$ & $C_R^{\mathrm{Ch}}/D_n$ & $C_D^{\mathrm{Ch}}/D_n$ & Coverage R & Coverage D\\\midrule
1,000 & 0.947063 & 0.3806 & 0.4627 & 0.9988 & 1.0000\\
4,000 & 0.990000 & 0.2171 & 0.2349 & 0.9992 & 0.9999\\
16,000 & 0.997500 & 0.1179 & 0.1225 & 0.9996 & 0.9999\\
64,000 & 0.999375 & 0.0623 & 0.0635 & 0.9999 & 0.9999\\
\bottomrule\end{tabular}
\caption{Comparison at $G_-$. The lower bound concerns deterministic rules honest over the unrestricted-probability reliability class. R is randomised and honest over the same class. D is deterministic and honest over the smaller reference-interior subclass. All columns use the same clean generating law. R and D report simulated mean diameter divided by $D_n$, and coverage integrates the seed.}\label{supp-tab:samelaw}
\end{table}
Both procedures have empirical coverage near one at $G_-$. Across the four sizes, the observed $\E\diam C_D^{\mathrm{Ch}}/\sqrt n$ values are $7.32$, $7.43$, $7.75$ and $8.04$. This favourable-law comparison complements the uniform $O(\sqrt{n\log n})$ guarantee in Theorem~\ref{thm:interior}.

At declared $\eta=0.05$, diameter distributions are evaluated at $G_-$ and the two-state model (Table~\ref{supp-tab:diameter_distribution}). For state weights $w_u$, seed-branch probabilities $f_u$, and empirical base-diameter CDFs $F_u$, the CDF is
\[
F(x)=\sum_u w_u\{f_u\mathbf1\{D_n\le x\}+(1-f_u)F_u(x)\}.
\]
Quantiles are $\inf\{x:F(x)\ge p\}$. This mixes the full-range atom with each base-diameter distribution before taking quantiles. The full-range probability also includes base sets whose diameter is $D_n$. The mean agrees with the earlier state-weighted results. Let $\widehat e_u$ denote the simulated proportion of base sets whose lower endpoint exceeds their upper endpoint. The estimated empty-set probability is
\[
\widehat p_{\varnothing}=\sum_u w_u(1-f_u)\widehat e_u.
\]
Empty sets contribute zero diameter and count as coverage failures.
\begin{table}[H]\centering\small
\begin{tabular}{lrlrrrrr}\toprule
Law & $n=m$ & Rule & $\Pp(\diam C=D_n)$ & $\Pp(C=\varnothing)$ & Median$/D_n$ & $q_{0.9}/D_n$ & Mean$/D_n$\\\midrule
$G_-$ & 4,000 & D & 0.1309 & 0.000017 & 0.1255 & 1.0000 & 0.2349\\
$G_-$ & 4,000 & R & 0.1306 & 0.000504 & 0.1050 & 1.0000 & 0.2171\\
$G_-$ & 64,000 & D & 0.0362 & 0.000039 & 0.0287 & 0.0394 & 0.0635\\
$G_-$ & 64,000 & R & 0.0362 & 0.000058 & 0.0274 & 0.0381 & 0.0623\\
Two-state & 4,000 & D & 0.1309 & 0.043457 & 0.2945 & 1.0000 & 0.3732\\
Two-state & 4,000 & R & 0.1306 & 0.043470 & 0.2330 & 1.0000 & 0.3225\\
Two-state & 64,000 & D & 0.0362 & 0.048190 & 0.0695 & 0.0703 & 0.0999\\
Two-state & 64,000 & R & 0.0362 & 0.048190 & 0.0648 & 0.0655 & 0.0955\\
\bottomrule\end{tabular}
\caption{Diameter distributions with declared $\eta=\alpha=0.05$. Actual failure probability is zero at $G_-$ and $0.05$ in the two-state model. The full-range branch has probability above $0.1$ at $n=m=4{,}000$, so the 90th percentile is $D_n$. Probability columns are simulation estimates with seed and state probabilities integrated. Means, medians and quantiles use the same mixed distribution.}\label{supp-tab:diameter_distribution}
\end{table}

\subsection{Fixed linear maps}\label{supp-sec:certificates}

The certified reserve lengths use $\tau=0.1$, $d=\lceil\log_2m\rceil$ and target $\Delta_A\le1/(2m)$ (Table~\ref{supp-tab:cert}). The simulations use these matrices and the threshold bound $\Delta_{\rm seed}=1/(2m)$. Each matrix is fixed before data collection.

The four binary matrices and the simulation scripts are supplied in \path{reproduction/matrices/} and \path{reproduction/code/}, respectively, in the reproduction archive available from the authors, with execution instructions in \path{reproduction/README.md}.
\begin{table}[H]\centering\small
\begin{tabular}{rrrrr}\toprule
$m$ & Bits $d$ & Linear $h$ & Linear $h/m$ & TV bound\\\midrule
1,000 & 10 & 105 & 0.1050 & $4.77\times10^{-4}$\\
4,000 & 12 & 126 & 0.0315 & $1.18\times10^{-4}$\\
16,000 & 14 & 146 & 0.0091 & $3.01\times10^{-5}$\\
64,000 & 16 & 168 & 0.0026 & $7.27\times10^{-6}$\\
\bottomrule\end{tabular}
\caption{Reference reserve and uniform seed-error bounds for the fixed linear maps. The TV column gives the uniform bound from Lemma~\ref{supp-lem:linear}.}\label{supp-tab:cert}
\end{table}
\Needspace{7\baselineskip}
Let $a_w$ count nonzero codewords of weight $w$. At $\tau=0.1$, $\rho=4/5$, so the certificate has the following integer form.
\[
\Delta_A\le\frac1{2m}
\quad\Longleftrightarrow\quad
m^2\sum_{w=0}^h a_w16^w25^{h-w}\le25^h.
\]
All four matrices satisfy this inequality. Their minimum nonzero weights are 36, 41, 51 and 58, respectively. These certificates hold for every allowed conditional probability vector.

\FloatBarrier

\clearpage
\begingroup\small\raggedright
\interlinepenalty=10000
\bibliographystyle{unsrt}
\bibliography{main}
\endgroup
\end{document}